\documentclass[arxiv]{melba}
\usepackage{amsmath,amsfonts}

\usepackage{booktabs}   % For professional looking tables
\usepackage{graphicx}   % To resize the table to fit page width
\usepackage{subcaption}
\usepackage{commath}
\usepackage{xurl}
\usepackage{stfloats}

\melbaid{YYYY:NNN}  % This is provided upon by the publishing editor
\doi{10.59275/j.melba.2024-AAAA}
\melbaauthors{Hoareau et al.}  % Note: this one is also used to set the pdf 'authors' metadata
\email{paul.hoareau@etud.polymtl.ca}
\volume{3}
\firstpageno{1337}  % Communicated by the publishing editor
\melbayear{YYYY}  % The publication year
\datesubmitted{yyyy-m1-d1}  % Date submitted to MELBA: mm/yyyy
\datepublished{yyyy-m2-d2}  % Today's date: mm/yyyy

\ShortHeadings{Weakly Supervised Segmentation of Ex Vivo MRI}{Hoareau et al.}

\title{Optimization in Sparse 2D to Dense 3D Weakly Supervised Learning: Application to Multi-Label Segmentation of Large ex vivo MRI Data}

\author{
	\firstname Paul \surname Hoareau\aff{1, 2, 3},
	\firstname Kuan Yi \surname Wang\aff{1},
    \firstname Brandon \surname Bujak\aff{11},
    \firstname Roy \surname Sun\aff{11},
    \firstname Govind \surname Nair\aff{11},
    \firstname Irene \surname Cortese\aff{9,12},
    \firstname Charidimos \surname Tsagkas\aff{5,6,7,8},
    \firstname Daniel \surname Reich\aff{9},
    \firstname Julien \surname Cohen-Adad\aff{1,3,10}
}
\affiliations{% <- trailing '%' to avoid unwanted indent
	\num 1 \addr NeuroPoly Lab, Institute of Biomedical Engineering, Polytechnique Montreal, Montreal, QC, Canada \\
    \num 2 \addr École Centrale de Lyon, Lyon, France \\
    \num 3 \addr Mila - Quebec AI Institute, Montreal, QC, Canada \\
	\num 4 \addr Functional Neuroimaging Unit, CRIUGM, University of Montreal, Montreal, QC, Canada \\
    \num 5 \addr Translational Neuroradiology Section, National Institute of Neurological Disorders and Stroke, National Institutes of Health, 10 Center Dr, Bethesda, MD 20893, USA \\
    \num 6 \addr Translational Imaging in Neurology (ThINk) Basel, Department of Biomedical Engineering, Faculty of Medicine, University Hospital Basel and University of Basel, Switzerland \\
    \num 7 \addr Neurologic Clinic and Policlinic, Departments of Medicine, University Hospital Basel, Switzerland \\
    \num 8 \addr Research Center for Clinical Neuroimmunology and Neuroscience Basel (RC2NB), University Hospital Basel and University of Basel, Switzerland \\
    \num 9 \addr National Institute of Neurological Disorders and Stroke, National Institutes of Health, Bethesda, MD, USA \\
    \num 10 \addr Centre de recherche du CHU Sainte-Justine, Université de Montréal, Montreal, QC, Canada \\
    \num 11 \addr Quantitative MRI core facility, NINDS, NIH \\
    \num 12 \addr Experimental Immunotherapeutics Unit, Division of Neuroimmunology and Neurovirology, NINDS, NIH
}

\abstract{%   <- trailing '%' for backward compatibility of .sty file
        INTRODUCTION | The fully supervised segmentation of 3D medical images featuring large amounts of slices (1,000+), such as serial block-face electron microscopy, or high-resolution ex vivo MRI, imposes prohibitive cost of volumetric manual annotation. Oftentimes, researchers resort to limited ground truth annotations consisting of sparse 2D slices, forcing a reliance on models that lack volumetric context. Incorporating across-slice information in the model can be done using a weakly supervised Sparse-to-Dense framework. However, guidelines for training such models remain ambiguous, specifically regarding the alignment between human-centric visual enhancements and machine perception, and the transferability of optimization strategies from 2D to 3D contexts. Here, we propose an analysis of divergent regularization needs in the context of multi-class segmentation of high-resolution ex vivo spinal cord MRI. \\
        METHODS | Data consists of 9.4T MRI of ex vivo spinal cord volumes with sparse manual annotations (428 annotated slices). Samples were obtained from pathologically confirmed cases of Multiple Sclerosis, totaling 104,000+ slices. A 2D Teacher network was trained on those sparse annotations to generate dense volumetric pseudo-labels, which were subsequently used to train a 3D Student network. We conducted a systematic optimization of this pipeline, analyzing the impact of image preprocessing (to evaluate human-centric image contrast enhancement), spatial augmentation (to mitigate positional shortcut learning), and soft-label regularization (to model lesion boundary uncertainty) on both 2D and 3D architectures. \\
        RESULTS | Our analysis reveals a critical divergence in training dynamics between dimensions that challenges standard low-data training assumptions. While the 2D Teacher significantly benefited from strong spatial augmentation and soft-labeling regularization – improving White Matter Lesion detection by over 11 Dice Score percentage points – these same techniques failed to improve or even degraded the performance of the 3D Student. Additionally, preprocessing techniques designed to enhance visual contrast for human raters (such as CLAHE) caused a performance drop, degrading Gray Matter Lesion Dice Scores by nearly 25 percentage points by disrupting global statistical cues. \\
        DISCUSSION | This study sheds light on a perception divergence: preprocessing methods designed to aid human raters (image contrast enhancement) proved destructive to machine learning models by distorting global intensity statistics. Moreover, this study reveals a regularization conflict across dimensions in sparse-to-dense learning. While aggressive spatial augmentation and soft-labeling were helpful for the 2D Teacher to overcome sparse data scarcity, propagating these distinct techniques to the 3D Student resulted in performance degradation. These findings suggest that 3D architectures, when trained on dense pseudo-labels, exhibit fundamentally different optimization landscapes than their 2D counterparts and require a distinct, conservative regularization strategy. %
        Code and model are available at~\url{https://github.com/ivadomed/model_seg_sc-gm-lesion_human_ms_exvivo_t2star}.}

\keywords{Spinal Cord, Multiple Sclerosis, MRI, High Resolution, Segmentation}

\begin{document}

% top matter
% \twocolumn[\maketitle]
% comment the preceedings and uncomment the following if the authors list + abstract is longer than one page
\maketitle
\twocolumn

% Introduction (or first section)
% \rule{\textwidth}{1pt}
\section{Introduction}
	\enluminure{M}{ultiple} sclerosis pathology in the spinal cord is a primary driver of physical disability, frequently leading to significant impairments in ambulation and coordination \citep{Kearney2015-th, Kreiter2024}. While spinal cord white matter lesions are widely studied, histopathological assessments reveal that the spinal cord gray matter is consistently and extensively affected by demyelination \citep{Waldman2024, Kreiter2024}. In fact, gray matter involvement is highly prevalent and increasingly associated with progressive disease phenotypes \citep{Kreiter2024}. However, despite their clinical relevance, spinal cord gray matter lesions remain notoriously difficult to visualize and segment in vivo. Imaging the spinal cord inherently suffers from unique challenges, including its small cross-sectional size, physiological motion from proximity to moving organs, and partial volume effects that create ambiguous tissue boundaries \citep{NagaKarthik2025}. These physical limitations make the detection and segmentation of small internal targets, such as gray matter lesions, exceptionally cumbersome and prone to high inter-rater variability \citep{NagaKarthik2025}.

High-resolution ex vivo MRI provides a detailed anatomical characterization of both underlying anatomical structures and focal lesions. However, this high resolution (typically 50-100 µm isotropic) presents a significant computational and logistical challenge: a single ex vivo sample can contain thousands of slices. Creating a dense, voxel-wise manual ground truth for such volumes is impractical. Consequently, researchers are often forced to rely on sparse annotations (segmenting only a few disconnected slices per volume).

The lack of 3D ground truths creates a technical dichotomy. 2D models can be trained on sparse slices but inherently lack volumetric context, often producing predictions that are inconsistent along the Z-axis. Conversely, 3D models require dense volumetric ground truth which does not exist.

In this work, we propose a Sparse-to-Dense framework tailored for high-resolution ex vivo spinal cord MRI. Our approach leverages sparse 2D annotations to train a 2D teacher model, which generates dense pseudo-labels to train a 3D student model. We make the following contributions:
    \begin{itemize}
        \item \textbf{The Perception Divergence:} We provide empirical evidence that preprocessing techniques optimized for human perception (CLAHE, contrast stretching) effectively impair machine performance in MRI segmentation, degrading Gray Matter Lesion (Lesion GM) Dice Score by up to 25 percentage points.
        \item \textbf{Divergent Regularization Needs:} We identify a critical conflict in Sparse-to-Dense pipelines: while 2D models benefit from aggressive regularization (spatial augmentation, soft labels) to prevent overfitting, these exact same techniques degrade the performance of the 3D model.
        \item \textbf{Open access segmentation model:} We train a model for segmenting white matter, gray matter and multiple sclerosis lesions on ex vivo human spinal cord MRI and make it available for the community via the open source software Spinal Cord Toolbox  \citep{De_Leener2017-um}.
    \end{itemize}

%%%%%%%%%%%%%%%%%%%%%%%%%%%%%%%%%%%%%%%%%%%%%%%%%%%%%%%%%%%%%%%%%%%%%%%%%%%
% Related works
%%%%%%%%%%%%%%%%%%%%%%%%%%%%%%%%%%%%%%%%%%%%%%%%%%%%%%%%%%%%%%%%%%%%%%%%%%%
% Make sure to put your work into context and include apporpriate citations.
% We do not have limits on citation counts.
\section{Related Works}
\subsection{The Challenge of Sparse-to-Dense in Medical Imaging}
The development of robust volumetric segmentation models in medical imaging is frequently obstructed by the annotation bottleneck. Creating a dense, voxel-wise manual ground truth for high-resolution datasets is a labor-intensive task that requires significant effort from domain experts, making it prohibitively expensive to obtain in clinical scenarios \citep{Hesamian2019-gn}. As \cite{Tajbakhsh2020-kt} note, rarely do researchers have access to a ``perfect'' training dataset; instead, they are often forced to rely on ``scarce annotations,'' where only a limited number of labeled scans are available, or ``weak annotations,'' where the training data has sparse annotations, noisy labels, or image-level tags \citep{Hesamian2019-gn}.

\subsubsection*{The ``Dimension Gap'': 2D Efficiency vs. 3D Consistency}

The scarcity of supervision signals forces a technical dichotomy in model development often referred to as the ``Dimension Gap'':
\begin{itemize}
    \item \textbf{Option A:} 2D Models. Standard 2D Convolutional Neural Networks (CNNs) are highly data-efficient and compatible with sparse slice annotations. However, because they treat each slice independently, they inherently lack volumetric context \citep{Hesamian2019-gn}. This limitation often leads to inconsistent predictions along the longitudinal axis, failing to capture inter-slice information. To mitigate this, some approaches employ ``2.5D'' strategies, extracting orthogonal patches (e.g., XY, YZ, XZ planes) to incorporate richer spatial information without the full computational cost of 3D networks.
    \item \textbf{Option B:} 3D Models. 3D CNNs theoretically offer superior performance by extracting volumetric representations across all three axes and utilizing 3D max-pooling to stabilize learned features \citep{Hesamian2019-gn}. However, they are computationally demanding and prone to overfitting when trained on limited data due to the ``curse of dimensionality'' and the massive number of parameters required. Furthermore, training these models typically requires dense volumetric ground truth, which is exactly what is missing in sparse annotation scenarios \citep{Cai2023-mh}.
\end{itemize}

Work by \cite{Perslev2019-cp} challenges this strict dichotomy, demonstrating that a ``Multi-Planar U-Net'' (MPUnet) (a 2D architecture trained with extensive multi-view augmentation) can learn a representation of the 3D volume that fosters generalization without the heavy computational footprint of native 3D models. By fusing predictions from multiple fixed planes, MPUnet avoids the heavy computational footprint of native 3D models while maintaining spatial consistency. This highlights that bridging the dimension gap is often a matter of how the data is presented to the network, not just the network architecture itself.

\subsubsection*{Existing Solutions: Pseudo-Labeling and Cross-Teaching}

A particularly relevant evolution of this concept is the ``3D-2D Cross-Teaching'' paradigm proposed by \cite{Cai2023-mh}. In this framework, 2D and 3D networks co-train each other: the 2D networks leverage their slice-wise data efficiency to supervise a 3D network, which in turn regularizes the 2D predictions with volumetric context. Unlike traditional registration-based methods that struggle with anatomical variance \citep{Cai2023-mh}, this approach imposes consistency between dimensions. This aligns with semi-supervised co-training approaches, where networks generate credible pseudo-labels for one another, effectively transferring the student's focus from sparse slices to full volumetric consistency \citep{Cai2023-mh}.

\subsection{Preprocessing \& Image Contrast Enhancement}
In medical image analysis, a fundamental tension exists between preprocessing that optimizes for human perception (visual clarity) and that which preserves statistical consistency for machine learning. Contrast Limited Adaptive Histogram Equalization (CLAHE) is the standard for the former, developed to reveal structures hidden by wide dynamic ranges for human observers \citep{Zuiderveld1994-nk}. While modern frameworks like nnU-Net typically discourage such ``superfluous bells and whistles'' in favor of dynamic network adaptation \citep{Isensee2021-bc}, recent evidence suggests image contrast enhancement can still offer specific benefits. For instance, \cite{Yoshimi2024-gj} found that CLAHE improved the robustness of segmentation models against domain shifts when testing on unseen data from new scanners, although it offered no statistically significant advantage when the training data was already heterogeneous.

Driven by the visual difficulty of our specific task, we hypothesized that baking these visual enhancements into the training tensors would reduce the burden on the model to learn complex non-linear intensity mappings. During the ground truth generation process, manual raters heavily relied on dynamic windowing to visualize low-contrast gray matter boundaries that were otherwise imperceptible in the raw linear data. To mimic this human-guided feature extraction, we investigated two targeted interventions: Phase Contrast Stretching - rather than CLAHE - to directly address the natively low contrast of the phase signal and replicate the manual windowing used to distinguish gray from white matter. Simultaneously, we applied Magnitude CLAHE to highlight local texture details within lesion boundaries. This design directly tests whether the ``visual clarity'' required for manual rating translates to improved latent feature extraction for CNNs.

\subsection{Shortcut Learning \& Geometric Augmentation}
Convolutional Neural Networks (CNNs) are prone to ``shortcut learning'' \citep{Geirhos2020-an}, a phenomenon where the model solves a task by relying on spurious correlations - such as background statistics or positional cues - rather than learning intrinsic semantic features. This risk is particularly acute in our standardized spinal cord dataset, where the anatomy is consistently centered. Without intervention, the network may learn to identify structures based solely on their spatial coordinates (e.g., ``the center usually contains gray matter'') rather than their biological shape. This texture and position bias is well-documented; \cite{Geirhos2018-cc} demonstrated that standard CNNs often fail to recognize objects when texture cues are removed, even if the shape is preserved.

To counteract these biases, we adopted a strategy of aggressive geometric data augmentation. \cite{Chaitanya2020-ae} reported that extensive spatial augmentation serves as a formidable baseline in medical imaging with limited annotations, often rivaling complex semi-supervised methods. Furthermore, as demonstrated by \cite{Oza2022-ba} in mammography and \cite{Isensee2021-bc} with nnU-Net, ``unrealistic'' deformations (such as heavy rotations or twisting) are highly effective drivers of generalization. These transformations break the spurious correlations between position and anatomy, compelling the network to learn shape-invariant features robust to the orientation or location of the spinal cord.

\subsection{Boundary Uncertainty \& Soft Labels}

Traditional segmentation loss functions, such as Dice and cross-entropy, treat anatomical boundaries as absolute binary truths (0 or 1). However, this binary approach often conflicts with medical reality where in tissue structures can lack sharp boundaries, as opposed to objects in natural images. \cite{Commowick2018-cv} demonstrated that manual annotation is inherently subjective; even expert consensus is difficult to obtain, suggesting that penalizing a model for misclassifying pixels in these ambiguous transition zones can be counter-productive.

To address this, we hypothesize that adopting a ``Boundary Uncertainty'' strategy, as proposed by \cite{Yeung2021-di} can improve generalization by mimicking human uncertainty. By dynamically assigning probabilistic scores to pixels at the structural edges, the model is permitted to be less confident in transition zones while remaining highly confident in the core of the tissue. However, the application of soft labels requires careful calibration \citep{Gros2021-gx}. As \cite{Muller2019-lo} noted, while label smoothing improves calibration, it can encourage representations to cluster too tightly, potentially erasing subtle semantic details. Therefore, our investigation focuses on whether spatially-restricted soft labeling acts as a beneficial regularizer for spinal cord anatomy or if it leads to destructive over-smoothing.

% A methodological, model, or similar section often comes here.
\section{Methods}
\subsection{Dataset and Acquisition}

The study utilizes a dataset of 12 ex vivo spinal cord samples. To foster robust feature extraction of diverse T2*-hyperintense pathological tissues, the training and validation cohorts encompass samples from patients with multiple sclerosis as well as progressive multifocal leukoencephalopathy. However, to ensure a precise and targeted evaluation aligned with our primary objective, the held-out test set consists exclusively of multiple sclerosis cases. The samples were scanned intact in consecutive segments (``chunks'') featuring a 5.5 cm rostrocaudal length and a 0.5 cm overlap. Imaging was conducted on a 9.4 T Bruker (Biospec 9.4 T/30 cm) pre-clinical scanner with an 86 mm transmit-receive volume coil. We acquired T2*-weighted images at a 75 µm isotropic resolution (TR: 40 ms; TE: 9 ms; flip angle = 44°; matrix: 200 × 306 × 730; 12–21 averages; 45.9 min/average), generating 43,719 axial slices. Both magnitude and phase images (left-handed convention, paramagnetic = bright) were utilized for training the segmentation model. 

\subsection{Data Splitting}

To ensure rigorous validation, the dataset was split at the case level rather than the chunk level. This acts as a necessary precaution against plausible data leakage, as chunks from the same spinal cord could share inherent similarities in anatomical geometry and lesion distribution.

\subsection{Ground Truth and Class Definitions}

\begin{figure*}
    \centering
    \includegraphics[width=1\linewidth]{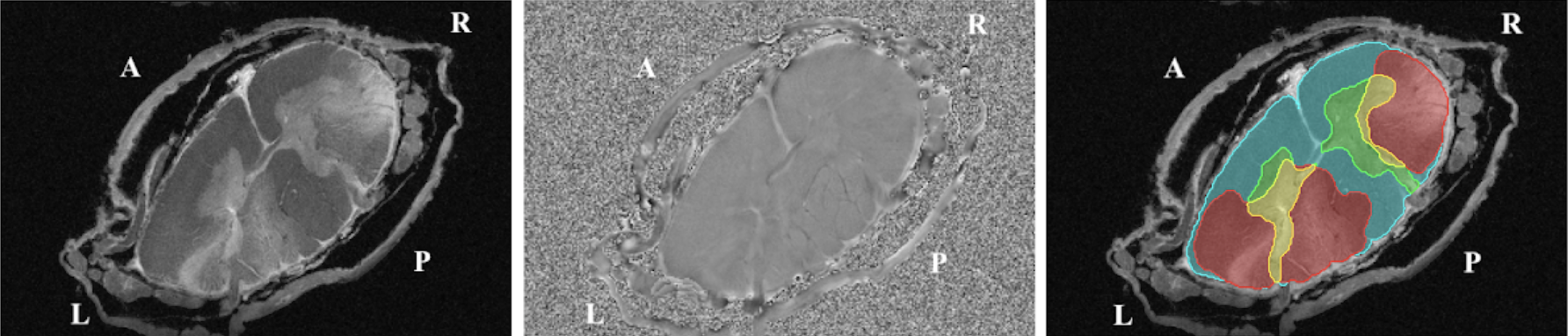}
    \caption{Illustration of the magnitude, the phase and the associated segmentation mask. Cyan: Healthy WM; Green: Healthy GM; Red: Lesion WM; Yellow: Lesion GM. All images in this manuscript are not standardized but shown exactly as inputted to the models to preserve objective visual representation.}
    \label{fig:illus_mag_phase}
\end{figure*}

Due to the high cost of volumetric annotation at this resolution, we employed a sparse annotation strategy. A total of 428 axial slices were manually segmented by PH and verified by a neuroradiologist with more than 10 years of neuroimaging experience in multiple sclerosis (CT)  across the training set. The segmentation task comprises four semantic classes as illustrated in Figure~\ref{fig:illus_mag_phase}:
\begin{itemize}
    \item Healthy White Matter (Healthy WM)
    \item Healthy Gray Matter (Healthy GM)
    \item Lesion in White Matter (Lesion WM)
    \item Lesion in Gray Matter (Lesion GM)
\end{itemize}

\subsection{Model Framework}

We utilized nnU-Net v2 \citep{Isensee2021-bc}, a self configuring U-Net variant that automatically adapts its architecture to the dataset fingerprint. To conduct our experiments we create a custom trainer and a customized inference code. All code files are available on this project GitHub repository at~\url{https://github.com/ivadomed/model_seg_sc-gm-lesion_human_ms_exvivo_t2star}. 

\textbf{Region-Based Training:} To address the structured nature of spinal cord anatomy (where lesions spatially reside within tissues) we employed region-based training.  Rather than treating the four classes as mutually exclusive during loss calculation, the model predicts overlapping regions: ``white matter,'' ``gray matter,'' and ``all lesions.'' These regional predictions are subsequently merged to resolve the final four mutually exclusive labels (e.g., Lesion WM is the intersection of the white matter and all lesion predictions).

\subsection{The Sparse-to-Dense Training Pipeline}

Our framework bridges the gap between 2D sparse labels and 3D inference through a three-stage process:

Stage 1: The 2D Teacher | A 2D nnU-Net is trained on the 374 sparsely annotated slices (the 54 remaining slices are held out as the test set). As detailed in Appendix~\ref{app:strategy}, we performed an extensive investigation to determine the optimal input configuration and training dynamics for this teacher model.

Stage 2: Dense Pseudo-Label Generation | The best-performing 2D configuration was used to run inference on the entirety of the training volumes (64 volumes, 12 cases, 43,719 slices). To generate the highest quality pseudo-ground truth, we employed ensembling (combining predictions from 4-fold cross-validation models) and Test Time Augmentation (TTA). Note on Label Fidelity: For this specific generation step, we utilized the models on the data they were trained on (conceptually similar to overfitting). This ensures that the propagated pseudo-labels are as accurate as possible to the available manual ground truth, this is also natural when doing ensembling.

Stage 3: The 3D Student | A 3D nnU-Net model is trained using the dense pseudo-labels generated in Stage 2. This allows the model to learn 3D volumetric features and longitudinal consistency that were inaccessible to the 2D model. We conducted a similar extensive investigation to determine the optimal input configuration and training dynamics for this student model.

\subsection{Evaluation Methodology}

To strictly evaluate model performance while accounting for biological variability, we employed a 4-fold cross-validation strategy respecting case independence. Performance was assessed using two primary approaches:

Slice-wise Accuracy (Validation \& Test): We calculated the Dice Similarity Coefficient (Dice Score) for each class on the 374 sparsely annotated axial slices in the cross validation folds and the 54 annotated slices of the held-out test set. The Dice Score serves as our primary accuracy metric against manual ground truth and allows us to get a global understanding of how the model performs on each class:

\begin{equation}
\text{Dice Score}(G, P) = \frac{2|G \cap P|}{|G| + |P|}
\label{eq:Dice Score}
\end{equation}

We report the Dice Score independently for each semantic class to analyze performance on structures of varying sizes, as well as the mean Dice across all foreground classes. We evaluate both the 2D models and the 3D models using the manually annotated slices only. 

95th Percentile Hausdorff Distance (HD95): While the Dice Score quantifies volumetric overlap, it can be insensitive to boundary accuracy, particularly for complex shapes or small structures where slight spatial shifts result in disproportionate penalties. To address this, we employed HD95 to evaluate the spatial consistency of the segmentation boundaries. The Hausdorff Distance measures the maximum distance from a point in one set to the nearest point in the other. To mitigate the impact of outliers (e.g., isolated pixel noise common in medical segmentation), we utilize the 95th percentile rather than the absolute maximum:

\begin{equation}
\text{HD95}(G, P) = \max(h_{95}(G, P), h_{95}(P, G))
\label{eq:hd95}
\end{equation}

Where $h_{95}(G, P)$ represents the 95th percentile of the distances from surface points in the ground truth G to the nearest surface points in the prediction P. A lower HD95 distance signifies better boundary alignment between the prediction and the ground truth, and it is measured in millimeters.  

Geometric Consistency (Inter-slice Dice): Standard volumetric metrics (Dice Score, HD95) quantify global overlap but are insensitive to high-frequency discontinuities along the slicing axis. To specifically evaluate the anatomical plausibility of the 3D reconstruction and quantify the ``z-axis jitter'' often associated with 2D slice-wise prediction stacking, we calculated the inter-slice Dice Coefficient Score ($DSC_{z}$).

For a given semantic class, $DSC_{z}$ measures the harmonic mean of the overlap between consecutive axial slices z and z+1. To avoid artificially inflating the score with background-to-background transitions, we restricted the calculation to the set of valid transitions Z, defined as pairs where the class is present in at least one of the two slices:

\begin{equation}
Z = \{ z \mid |S_z| + |S_{z+1}| > 0 \}
\label{eq:valid_transitions}
\end{equation}

The metric is formally defined as:

\begin{equation}
DSC_z = \frac{1}{|Z|} \sum_{z \in Z} \frac{2|S_z \cap S_{z+1}|}{|S_z| + |S_{z+1}|}
\label{eq:inter_slice_Dice}
\end{equation}

where \abs{Z} represents the total count of valid transitions. A higher $DSC_{z}$ indicates superior longitudinal smoothness and geometric continuity.

\section{Results}
\subsection{Utility of magnitude and/or phase for model training}

As magnitude and phase images carry different information (phase data represents more susceptibility-weighted information), we compared single-input vs. dual-input models as follows:
\begin{itemize}
    \item 2D magnitude + phase: Input tensor size (B,2,H,W), 3D: Input tensor size  (B,2,H,W,Z).
    \item 2D magnitude only: Input tensor size (B,1,H,W), 3D: Input tensor size (B,1,H,W,Z).
\end{itemize}

Table~\ref{tab:mag_phase_2D} shows results of the 2D models. The removal of phase data resulted in a slight decrease in global performance ($<$1 percentage point Dice Score decrease in Total Average). However, the impact on stability was significant for the most challenging class, Lesion GM. While the magnitude-only model performed adequately for general detection, the coefficient of variation for Lesion GM increased substantially (3.1\% with both magnitude and phase to 8.0\% without phase), and the mean Dice Score dropped from 0.665 to 0.641. For Healthy White Matter, the removal of phase data degraded boundary precision, increasing the HD95 from 0.78 mm to 0.92 mm. Interestingly, for Lesion GM, the HD95 slightly improved (decreased) from 0.58 mm to 0.56 mm in the magnitude-only model, which could indicate bias toward magnitude information during the annotation task.

\begin{table*}[t]
    \centering
    \caption{Magnitude and Phase 2D Experiment -- Dice and HD95 2D Results. Best results are highlighted in bold.}
    \label{tab:mag_phase_2D}
    \resizebox{\textwidth}{!}{%
        \begin{tabular}{lcccccccccc}
            \toprule
            & \multicolumn{2}{c}{\textbf{Healthy WM}} & \multicolumn{2}{c}{\textbf{Healthy GM}} & \multicolumn{2}{c}{\textbf{Lesion WM}} & \multicolumn{2}{c}{\textbf{Lesion GM}} & \multicolumn{2}{c}{\textbf{Total Average}} \\
            \cmidrule(lr){2-3} \cmidrule(lr){4-5} \cmidrule(lr){6-7} \cmidrule(lr){8-9} \cmidrule(lr){10-11}
            \textbf{Config} & \textbf{Dice} & \textbf{HD95} & \textbf{Dice} & \textbf{HD95} & \textbf{Dice} & \textbf{HD95} & \textbf{Dice} & \textbf{HD95} & \textbf{Dice} & \textbf{HD95} \\
            \midrule
            2D Mag+Phase & $\mathbf{0.848 \pm 0.087}$ & $\mathbf{0.78 \pm 0.35}$ & $0.819 \pm 0.067$ & $0.74 \pm 0.32$ & $0.729 \pm 0.040$ & $\mathbf{0.92 \pm 0.27}$ & $\mathbf{0.665 \pm 0.021}$ & $0.58 \pm 0.28$ & $\mathbf{0.766 \pm 0.041}$ & $\mathbf{0.76 \pm 0.28}$ \\
            2D Mag & $0.842 \pm 0.098$ & $0.92 \pm 0.39$ & $\mathbf{0.824 \pm 0.065}$ & $\mathbf{0.74 \pm 0.34}$ & $\mathbf{0.735 \pm 0.027}$ & $1.01 \pm 0.37$ & $0.641 \pm 0.051$ & $\mathbf{0.56 \pm 0.30}$ & $0.760 \pm 0.046$ & $0.81 \pm 0.32$ \\
            \bottomrule
        \end{tabular}%
    }
\end{table*}

Table~\ref{tab:mag_phase_3D} shows results of the 3D models. Removing the phase data caused a general degradation in performance, dropping the total average Dice from 0.772 to 0.760. The impact was observed in the Healthy GM, dropping from a Dice of 0.827 to 0.823, and Lesion GM, dropping from 0.673 to 0.656. Crucially, the 3D HD95 metrics highlight a loss of geometric precision without phase data. The HD95 for Healthy GM worsened from 0.93 mm to 1.07 mm, and for Lesion GM, it slightly increased from 0.96 mm to 0.97 mm. This confirms that the phase signal is essential for the 3D model to accurately resolve the fine spatial boundaries of gray matter structures.

\begin{table*}[t]
    \centering
    \caption{Magnitude and Phase 3D Experiment -- Dice and HD95 3D Results. Best results are highlighted in bold.}
    \label{tab:mag_phase_3D}
    \resizebox{\textwidth}{!}{%
        \begin{tabular}{lcccccccccc}
            \toprule
            & \multicolumn{2}{c}{\textbf{Healthy WM}} & \multicolumn{2}{c}{\textbf{Healthy GM}} & \multicolumn{2}{c}{\textbf{Lesion WM}} & \multicolumn{2}{c}{\textbf{Lesion GM}} & \multicolumn{2}{c}{\textbf{Total Average}} \\
            \cmidrule(lr){2-3} \cmidrule(lr){4-5} \cmidrule(lr){6-7} \cmidrule(lr){8-9} \cmidrule(lr){10-11}
            \textbf{Config} & \textbf{Dice} & \textbf{HD95} & \textbf{Dice} & \textbf{HD95} & \textbf{Dice} & \textbf{HD95} & \textbf{Dice} & \textbf{HD95} & \textbf{Dice} & \textbf{HD95} \\
            \midrule
            3D Mag+Phase & $\mathbf{0.838 \pm 0.106}$ & $\mathbf{0.72 \pm 0.14}$ & $\mathbf{0.827 \pm 0.091}$ & $\mathbf{0.93 \pm 0.21}$ & $\mathbf{0.750 \pm 0.096}$ & $\mathbf{1.38 \pm 1.06}$ & $\mathbf{0.673 \pm 0.113}$ & $\mathbf{0.96 \pm 0.60}$ & $\mathbf{0.772 \pm 0.077}$ & $\mathbf{1.00 \pm 0.28}$ \\
            3D Mag & $0.833 \pm 0.109$ & $0.75 \pm 0.16$ & $0.823 \pm 0.091$ & $1.07 \pm 0.24$ & $0.727 \pm 0.117$ & $1.66 \pm 1.33$ & $0.656 \pm 0.121$ & $0.97 \pm 0.67$ & $0.760 \pm 0.084$ & $1.11 \pm 0.39$ \\
            \bottomrule
        \end{tabular}%
    }
\end{table*}

\subsection{Impact of Histogram Normalization (CLAHE/Gamma)}

To test the hypothesis detailed in Section 2.2, we implemented two targeted preprocessing methods illustrated in Figure~\ref{fig:prepro}. They were applied dynamically (``on-the-fly'') before network ingestion:

\begin{figure}[htbp]
    \centering
    
    % --- Row 1 ---
    \begin{subfigure}[b]{0.48\columnwidth}
        \centering
        \includegraphics[width=\textwidth]{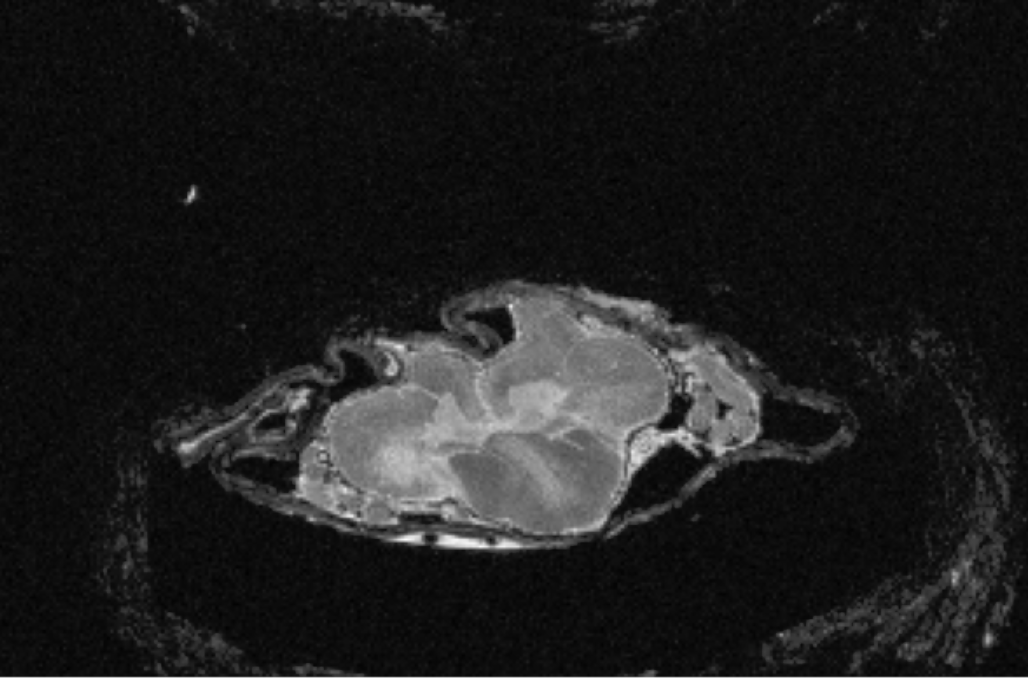} % Replace with your image file
        \caption{Original Magnitude}
        
        \label{fig:original}
    \end{subfigure}
    \hfill % Adds horizontal spacing between the two images in the row
    \hfill
    \begin{subfigure}[b]{0.48\columnwidth}
        \centering
        \includegraphics[width=\textwidth]{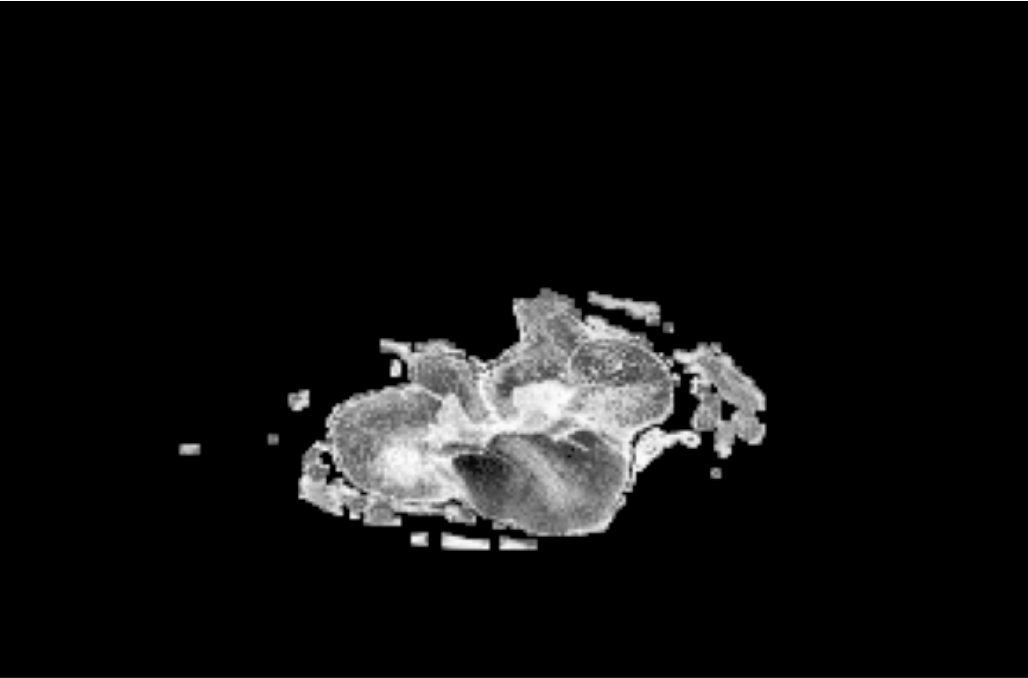}
        \caption{Preprocessed Magnitude}
        \label{fig:preprocessed}
    \end{subfigure}
    
    \vspace{1em} % Adds vertical space between the top and bottom rows
    
    % --- Row 2 ---
    \begin{subfigure}[b]{0.48\columnwidth}
        \centering
        \includegraphics[width=\textwidth]{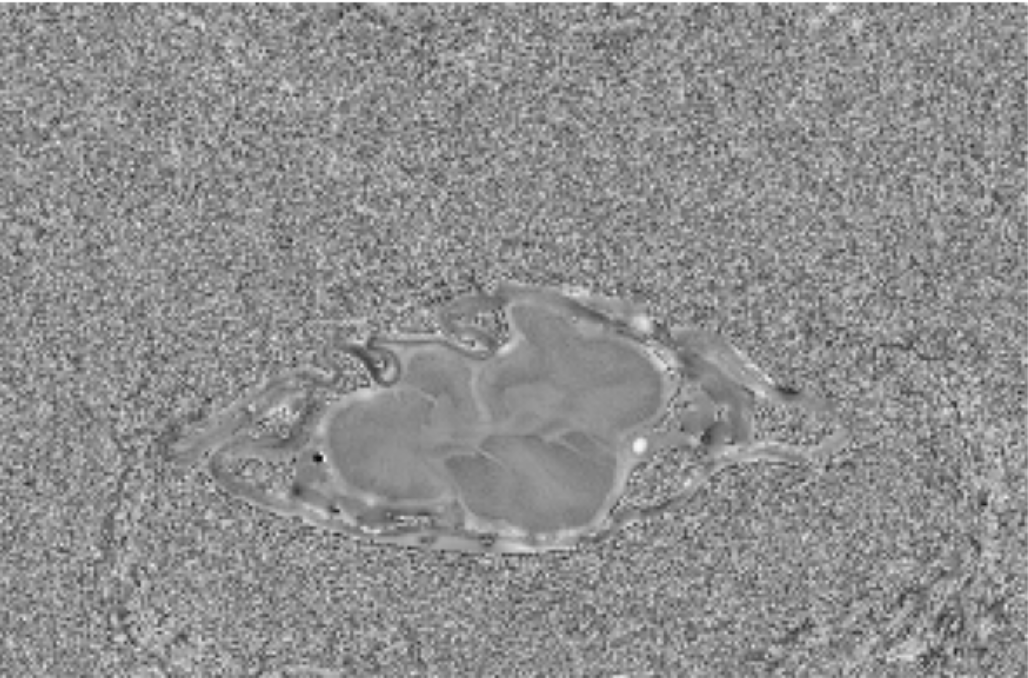}
        \caption{Original Phase}
        \label{fig:postprocessed}
    \end{subfigure}
    \hfill
    \begin{subfigure}[b]{0.48\columnwidth}
        \centering
        \includegraphics[width=\textwidth]{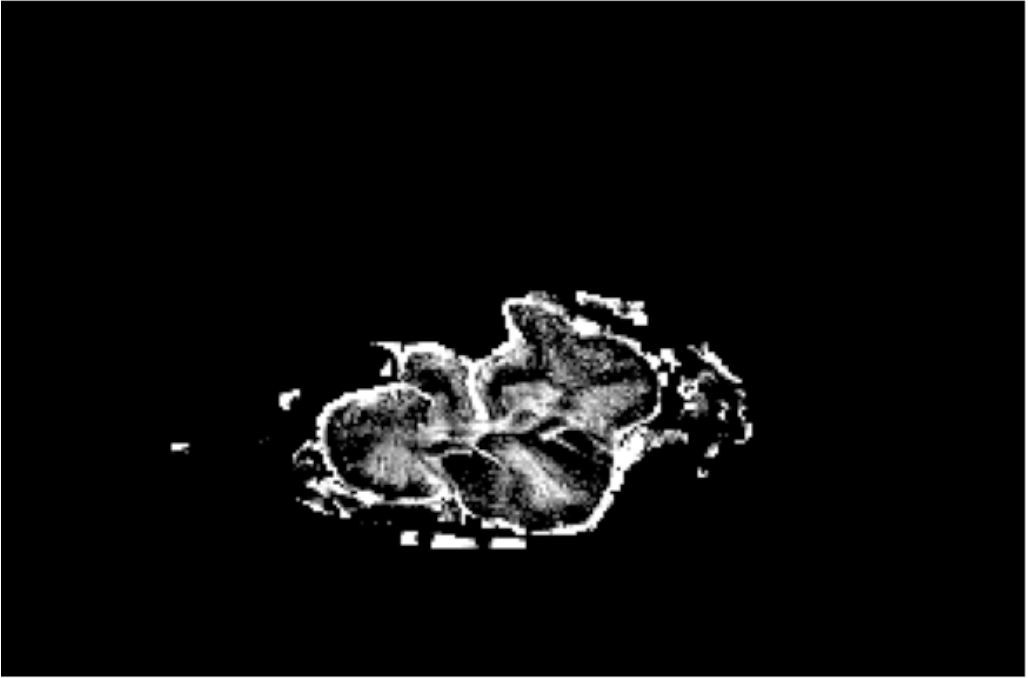}
        \caption{Preprocessed Phase}
        \label{fig:final}
    \end{subfigure}
    
    % --- Main Caption ---
    \caption{Illustration of the effect of our preprocessing. (a,c): Original Magnitude and Phase; (b): Magnitude + Otsu + CLAHE; (d)): Phase + Otsu + Gamma Stretching.}
    \label{fig:prepro}
\end{figure}

\begin{itemize}
    \item Magnitude CLAHE: We applied Contrast Limited Adaptive Histogram Equalization (CLAHE) to the T2*w magnitude channel to enhance local contrast in small regions (tiles).
    \item Phase Contrast Stretching: We applied percentile-based contrast stretching to the phase channel, clipping the bottom 15\% and top 30\% of intensities to maximize the dynamic range of the cord tissues.
\end{itemize}

Table~\ref{tab:prepro_2D} shows results of the 2D models. Contrary to expectations, both histogram normalization techniques were detrimental to model performance. As detailed in the table, Phase Contrast Stretching resulted in a moderate performance drop, reducing the Lesion GM Dice from 0.665 to 0.598. More drastically, Magnitude CLAHE caused a severe degradation, resulting in a Lesion GM Dice of 0.415 and a Lesion WM Dice of 0.490. The boundary metrics (HD95) corroborate this collapse in performance. Magnitude CLAHE caused the HD95 for Lesion WM to more than double (0.92 mm to 2.20 mm) and Lesion GM to spike from 0.58 mm to 1.25 mm. This indicates that the enhanced local contrast actively misled the model, causing it to predict lesions with wildly inaccurate and scattered boundaries.

\begin{table*}[t]
    \centering
    \caption{Preprocessing 2D Experiment -- Dice and HD95 2D Results. Best results are highlighted in bold.}
    \label{tab:prepro_2D}
    \resizebox{\textwidth}{!}{%
        \begin{tabular}{lcccccccccc}
            \toprule
            & \multicolumn{2}{c}{\textbf{Healthy WM}} & \multicolumn{2}{c}{\textbf{Healthy GM}} & \multicolumn{2}{c}{\textbf{Lesion WM}} & \multicolumn{2}{c}{\textbf{Lesion GM}} & \multicolumn{2}{c}{\textbf{Total Average}} \\
            \cmidrule(lr){2-3} \cmidrule(lr){4-5} \cmidrule(lr){6-7} \cmidrule(lr){8-9} \cmidrule(lr){10-11}
            \textbf{Config} & \textbf{Dice} & \textbf{HD95} & \textbf{Dice} & \textbf{HD95} & \textbf{Dice} & \textbf{HD95} & \textbf{Dice} & \textbf{HD95} & \textbf{Dice} & \textbf{HD95} \\
            \midrule
            2D Raw & $\mathbf{0.848 \pm 0.087}$ & $0.78 \pm 0.35$ & $\mathbf{0.819 \pm 0.067}$ & $\mathbf{0.74 \pm 0.32}$ & $\mathbf{0.729 \pm 0.040}$ & $\mathbf{0.92 \pm 0.27}$ & $\mathbf{0.665 \pm 0.021}$ & $\mathbf{0.58 \pm 0.28}$ & $\mathbf{0.766 \pm 0.041}$ & $\mathbf{0.76 \pm 0.28}$ \\
            2D Mag Prepro & $0.644 \pm 0.142$ & $1.10 \pm 0.38$ & $0.715 \pm 0.095$ & $1.15 \pm 0.51$ & $0.490 \pm 0.128$ & $2.20 \pm 0.64$ & $0.415 \pm 0.039$ & $1.25 \pm 0.35$ & $0.566 \pm 0.034$ & $1.43 \pm 0.18$ \\
            2D Phase Prepro & $0.821 \pm 0.096$ & $\mathbf{0.75 \pm 0.33}$ & $0.802 \pm 0.073$ & $0.80 \pm 0.34$ & $0.650 \pm 0.096$ & $1.04 \pm 0.29$ & $0.598 \pm 0.052$ & $0.67 \pm 0.28$ & $0.718 \pm 0.036$ & $0.82 \pm 0.29$ \\
            \bottomrule
        \end{tabular}%
    }
\end{table*}

Table~\ref{tab:prepro_3D} shows results of the 3D models. Similar to the 2D results, image contrast enhancement (applied to the magnitude and phase images) degraded performance compared to the Raw configuration. Phase preprocessing decreased the Dice Score for Lesion WM from 0.750 to 0.688, and magnitude preprocessing degraded the Lesion GM Dice from 0.673 to 0.635. Interestingly, while magnitude preprocessing degraded the Lesion GM Dice in 3D (0.673 to 0.635), it actually improved the Lesion GM HD95 (0.96 mm to 0.64 mm). This suggests the 3D model's volumetric context might offer some resilience against texture distortion regarding boundary placement, even if total overlap suffers.

\begin{table*}[t]
    \centering
    \caption{Preprocessing 3D Experiment -- Dice and HD95 3D Results. Best results are highlighted in bold.}
    \label{tab:prepro_3D}
    \resizebox{\textwidth}{!}{%
        \begin{tabular}{lcccccccccc}
            \toprule
            & \multicolumn{2}{c}{\textbf{Healthy WM}} & \multicolumn{2}{c}{\textbf{Healthy GM}} & \multicolumn{2}{c}{\textbf{Lesion WM}} & \multicolumn{2}{c}{\textbf{Lesion GM}} & \multicolumn{2}{c}{\textbf{Total Average}} \\
            \cmidrule(lr){2-3} \cmidrule(lr){4-5} \cmidrule(lr){6-7} \cmidrule(lr){8-9} \cmidrule(lr){10-11}
            \textbf{Config} & \textbf{Dice} & \textbf{HD95} & \textbf{Dice} & \textbf{HD95} & \textbf{Dice} & \textbf{HD95} & \textbf{Dice} & \textbf{HD95} & \textbf{Dice} & \textbf{HD95} \\
            \midrule
            3D Raw & $\mathbf{0.838 \pm 0.106}$ & $\mathbf{0.72 \pm 0.14}$ & $\mathbf{0.827 \pm 0.091}$ & $\mathbf{0.93 \pm 0.21}$ & $\mathbf{0.750 \pm 0.096}$ & $1.38 \pm 1.06$ & $\mathbf{0.673 \pm 0.113}$ & $0.96 \pm 0.60$ & $\mathbf{0.772 \pm 0.077}$ & $1.00 \pm 0.28$ \\
            3D Phase Prepro & $0.750 \pm 0.181$ & $0.85 \pm 0.24$ & $0.743 \pm 0.152$ & $1.06 \pm 0.49$ & $0.688 \pm 0.118$ & $1.23 \pm 0.90$ & $0.591 \pm 0.121$ & $0.80 \pm 0.51$ & $0.693 \pm 0.074$ & $0.99 \pm 0.20$ \\
            3D Mag Prepro & $0.801 \pm 0.121$ & $0.85 \pm 0.22$ & $0.793 \pm 0.103$ & $0.94 \pm 0.52$ & $0.689 \pm 0.104$ & $\mathbf{0.98 \pm 0.99}$ & $0.635 \pm 0.103$ & $\mathbf{0.64 \pm 0.72}$ & $0.729 \pm 0.081$ & $\mathbf{0.85 \pm 0.15}$ \\
            \bottomrule
        \end{tabular}%
    }
\end{table*}

\subsection{Data Augmentation}

To implement the shape-invariant learning strategy defined in Section 2.3, we applied three progressively aggressive augmentation profiles using on-the-fly spatial transformations. We utilized extensive geometric distortions including rotations up to 180°, aggressive scaling (0.1× to 3.0×), and heavy shearing (±85°). This unrealistic profile was designed to strictly penalize any reliance on the spinal cord's canonical centered position. All hyperparameters of all experiments are presented in Table~\ref{tab:spatial_aug_hyperparameters}.

\begin{table*}[htbp]
    \centering
    \caption{Spatial Augmentation Hyperparameters}
    \label{tab:spatial_aug_hyperparameters}
    \begin{tabular}{lccc}
        \hline
        \textbf{Transformation} & \textbf{Spatial Aug 1} & \textbf{Spatial Aug 2} & \textbf{Spatial Aug 3} \\
        \hline
        Translation & 0.45 & 0.45 & 0.80 \\
        Rotation ($^\circ$) & 90 & 180 & 180 \\
        Scale & (0.7, 1.7) & (0.3, 2.0) & (0.1, 3.0) \\
        Shear ($^\circ$) & (-35, 35) & (-55, 55) & (-85, 85) \\
        Perspective & 0.35 & 0.55 & 0.85 \\
        \hline
    \end{tabular}
\end{table*}

Table~\ref{tab:aug_2D} shows results of the 2D models. The omission of spatial augmentation proved highly detrimental. Without augmentation, the Lesion WM Dice degraded to 0.631 (vs. 0.729 for Aug1) and the HD95 increased to 1.51 mm (vs. 0.92 mm for Base), indicating a massive loss in boundary localization accuracy. Conversely, a moderately aggressive strategy (Spatial Aug 2) yielded the highest stability. It improved the Lesion WM Dice to 0.743 and achieved the best boundary precision for Lesion GM, lowering the HD95 to 0.53 mm (compared to 0.58 mm in Base). This confirms that heavy 2D augmentation is critical for teaching the model to localize boundaries accurately independent of the cord's position.

\begin{table*}[t]
    \centering
    \caption{Data Augmentation 2D Experiment -- Dice and HD95 2D Results. Best results are highlighted in bold.}
    \label{tab:aug_2D}
    \resizebox{\textwidth}{!}{%
        \begin{tabular}{lcccccccccc}
            \toprule
            & \multicolumn{2}{c}{\textbf{Healthy WM}} & \multicolumn{2}{c}{\textbf{Healthy GM}} & \multicolumn{2}{c}{\textbf{Lesion WM}} & \multicolumn{2}{c}{\textbf{Lesion GM}} & \multicolumn{2}{c}{\textbf{Total Average}} \\
            \cmidrule(lr){2-3} \cmidrule(lr){4-5} \cmidrule(lr){6-7} \cmidrule(lr){8-9} \cmidrule(lr){10-11}
            \textbf{Config} & \textbf{Dice} & \textbf{HD95} & \textbf{Dice} & \textbf{HD95} & \textbf{Dice} & \textbf{HD95} & \textbf{Dice} & \textbf{HD95} & \textbf{Dice} & \textbf{HD95} \\
            \midrule
            2D No Aug & $0.803 \pm 0.102$ & $\mathbf{0.69 \pm 0.27}$ & $0.786 \pm 0.087$ & $0.93 \pm 0.46$ & $0.631 \pm 0.084$ & $1.51 \pm 0.60$ & $0.603 \pm 0.005$ & $0.81 \pm 0.40$ & $0.706 \pm 0.043$ & $0.99 \pm 0.30$ \\
            2D Aug1 & $0.848 \pm 0.087$ & $0.78 \pm 0.35$ & $0.819 \pm 0.067$ & $0.74 \pm 0.32$ & $0.729 \pm 0.040$ & $0.92 \pm 0.27$ & $0.665 \pm 0.021$ & $0.58 \pm 0.28$ & $0.766 \pm 0.041$ & $0.76 \pm 0.28$ \\  
            2D Aug2 & $\mathbf{0.851 \pm 0.090}$ & $0.85 \pm 0.48$ & $\mathbf{0.825 \pm 0.064}$ & $\mathbf{0.73 \pm 0.32}$ & $\mathbf{0.743 \pm 0.033}$ & $\mathbf{0.84 \pm 0.34}$ & $\mathbf{0.667 \pm 0.021}$ & $\mathbf{0.53 \pm 0.23}$ & $\mathbf{0.772 \pm 0.042}$ & $\mathbf{0.74 \pm 0.33}$ \\
            2D Aug3 & $0.838 \pm 0.094$ & $0.87 \pm 0.46$ & $0.819 \pm 0.062$ & $0.79 \pm 0.34$ & $0.714 \pm 0.066$ & $0.90 \pm 0.33$ & $0.663 \pm 0.044$ & $0.56 \pm 0.26$ & $0.758 \pm 0.046$ & $0.78 \pm 0.33$ \\
            \bottomrule
        \end{tabular}%
    }
\end{table*}

Table~\ref{tab:aug_3D} shows results of the 3D models. We observe a divergence from the 2D stage. The No Spatial Augmentation approach achieved the highest performance across most metrics. The introduction of spatial augmentations (Aug 1, 2, 3) degraded the total average Dice (dropping from 0.772 to 0.724 in Aug 1) and significantly worsened the HD95 (increasing from 1.00 mm to 1.23 mm). The most aggressive 3D augmentation (Aug 3) caused a collapse in boundary fidelity, with the HD95 distance increasing to 1.28 mm.

\begin{table*}[t]
    \centering
    \caption{Data Augmentation 3D Experiment -- Dice and HD95 3D Results. Best results are highlighted in bold.}
    \label{tab:aug_3D}
    \resizebox{\textwidth}{!}{%
        \begin{tabular}{lcccccccccc}
            \toprule
            & \multicolumn{2}{c}{\textbf{Healthy WM}} & \multicolumn{2}{c}{\textbf{Healthy GM}} & \multicolumn{2}{c}{\textbf{Lesion WM}} & \multicolumn{2}{c}{\textbf{Lesion GM}} & \multicolumn{2}{c}{\textbf{Total Average}} \\
            \cmidrule(lr){2-3} \cmidrule(lr){4-5} \cmidrule(lr){6-7} \cmidrule(lr){8-9} \cmidrule(lr){10-11}
            \textbf{Config} & \textbf{Dice} & \textbf{HD95} & \textbf{Dice} & \textbf{HD95} & \textbf{Dice} & \textbf{HD95} & \textbf{Dice} & \textbf{HD95} & \textbf{Dice} & \textbf{HD95} \\
            \midrule
            3D No Aug & $\mathbf{0.838 \pm 0.106}$ & $\mathbf{0.72 \pm 0.14}$ & $\mathbf{0.827 \pm 0.091}$ & $\mathbf{0.93 \pm 0.21}$ & $\mathbf{0.750 \pm 0.096}$ & $\mathbf{1.38 \pm 1.06}$ & $\mathbf{0.673 \pm 0.113}$ & $\mathbf{0.96 \pm 0.60}$ & $\mathbf{0.772 \pm 0.077}$ & $\mathbf{1.00 \pm 0.28}$ \\
            3D Aug1 & $0.806 \pm 0.121$ & $0.94 \pm 0.19$ & $0.796 \pm 0.103$ & $1.11 \pm 0.21$ & $0.696 \pm 0.115$ & $1.78 \pm 1.17$ & $0.600 \pm 0.126$ & $1.08 \pm 0.62$ & $0.724 \pm 0.097$ & $1.23 \pm 0.38$ \\
            3D Aug2 & $0.772 \pm 0.134$ & $0.98 \pm 0.20$ & $0.663 \pm 0.153$ & $1.77 \pm 0.18$ & $0.508 \pm 0.143$ & $2.94 \pm 0.81$ & $0.415 \pm 0.106$ & $1.50 \pm 0.97$ & $0.590 \pm 0.159$ & $1.80 \pm 0.83$ \\
            3D Aug3 & $0.799 \pm 0.121$ & $0.97 \pm 0.21$ & $0.777 \pm 0.105$ & $1.20 \pm 0.32$ & $0.664 \pm 0.136$ & $1.75 \pm 1.10$ & $0.545 \pm 0.134$ & $1.19 \pm 0.76$ & $0.696 \pm 0.117$ & $1.28 \pm 0.33$ \\
            \bottomrule
        \end{tabular}%
    }
\end{table*}

\subsection{Hard vs. Soft Segmentation}

\begin{figure}
    \centering
    \scalebox{1}[-1]{\includegraphics[width=1\linewidth]{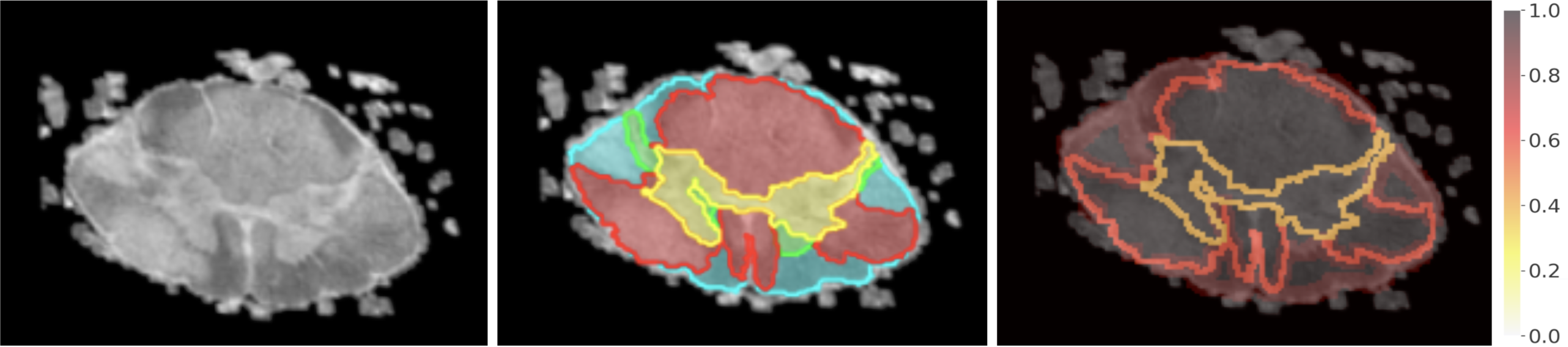}}
    \caption{Illustration of the soft edges on the magnitude. From left to right: magnitude, ground truth segmentation, soft edges map.}
    \label{fig:soft_edges}
\end{figure}

To operationalize the boundary uncertainty hypothesis defined in Section 2.4, we replaced standard binary targets with Soft Labels specifically within anatomical boundary zones, as illustrated by Figure~\ref{fig:soft_edges}. These zones were identified dynamically during training using morphological gradients (dilation minus erosion). Within these margins, we assigned soft probabilities ($\alpha$) determined by class-specific weights and kernel sizes presented in Table~\ref{tab:soft_seg_hyperparameters}. We created three distinct profiles:

\begin{itemize}
    \item Soft Seg 1: Applied moderate smoothing to lesion classes (Weights 0.4–0.6) over wide spatial margins (Kernel sizes 5–7).
    \item Soft Seg 2: Retained the moderate weights but tightened the spatial constraint on lesions (Kernel size 3) to prevent label bleeding into healthy tissues.
    \item Soft Seg 3: Tested extreme uncertainty by drastically reducing confidence for lesion targets (Weight 0.2) within tight margins.
\end{itemize}

\begin{table*}[htbp]
    \centering
    \caption{Soft Segmentation Hyperparameters}
    \label{tab:soft_seg_hyperparameters}
    \begin{tabular}{lcccccc}
        \hline
        & \multicolumn{2}{c}{\textbf{Soft Parameters 1}} & \multicolumn{2}{c}{\textbf{Soft Parameters 2}} & \multicolumn{2}{c}{\textbf{Soft Parameters 3}} \\
        \cline{2-7}
        \textbf{Class} & \textbf{Weight} & \textbf{Kernel Size} & \textbf{Weight} & \textbf{Kernel Size} & \textbf{Weight} & \textbf{Kernel Size} \\
        \hline
        Healthy WM & 0.9 & 7 & 0.9 & 7 & 0.7 & 5 \\
        Healthy GM  & 0.9 & 3 & 0.9 & 3 & 0.6 & 3 \\
        Lesion WM         & 0.6 & 5 & 0.6 & 3 & 0.2 & 3 \\
        Lesion GM         & 0.4 & 7 & 0.4 & 3 & 0.2 & 3 \\
        \hline
    \end{tabular}
\end{table*}

Table~\ref{tab:soft_2D} shows results of the 2D models. The standard hard training achieved a Lesion GM Dice of 0.665. We observed that applying aggressive smoothing specifically to the Lesion classes (Soft Loss 2) yielded the best performance, raising the Lesion WM Dice to 0.744 (+1.5 percentage points over Hard Labels) and the Lesion GM Dice to 0.672.
Soft Loss 2 achieved slightly tighter HD95 boundaries for the lesion classes: Lesion WM HD95 improved from 0.92 mm (Hard labels) to 0.91 mm, and Lesion GM HD95 improved from 0.58 mm to 0.56 mm. This suggests that soft labels successfully helped the 2D model refrain from over-penalizing pixels at the ambiguous lesion edges, resulting in contours that are geometrically closer to the ground truth.

\begin{table*}[t]
    \centering
    \caption{Soft Segmentation 2D Experiment -- Dice and HD95 2D Results. Best results are highlighted in bold.}
    \label{tab:soft_2D}
    \resizebox{\textwidth}{!}{%
        \begin{tabular}{lcccccccccc}
            \toprule
            & \multicolumn{2}{c}{\textbf{Healthy WM}} & \multicolumn{2}{c}{\textbf{Healthy GM}} & \multicolumn{2}{c}{\textbf{Lesion WM}} & \multicolumn{2}{c}{\textbf{Lesion GM}} & \multicolumn{2}{c}{\textbf{Total Average}} \\
            \cmidrule(lr){2-3} \cmidrule(lr){4-5} \cmidrule(lr){6-7} \cmidrule(lr){8-9} \cmidrule(lr){10-11}
            \textbf{Config} & \textbf{Dice} & \textbf{HD95} & \textbf{Dice} & \textbf{HD95} & \textbf{Dice} & \textbf{HD95} & \textbf{Dice} & \textbf{HD95} & \textbf{Dice} & \textbf{HD95} \\
            \midrule
            
            2D Hard & $\mathbf{0.848 \pm 0.087}$ & $0.78 \pm 0.35$ & $0.819 \pm 0.067$ & $\mathbf{0.74 \pm 0.32}$ & $0.729 \pm 0.040$ & $0.92 \pm 0.27$ & $0.665 \pm 0.021$ & $0.58 \pm 0.28$ & $0.766 \pm 0.041$ & $0.76 \pm 0.28$ \\
            2D Soft1 & $0.846 \pm 0.090$ & $\mathbf{0.71 \pm 0.30}$ & $0.823 \pm 0.066$ & $0.76 \pm 0.36$ & $0.725 \pm 0.043$ & $0.97 \pm 0.33$ & $0.664 \pm 0.023$ & $0.54 \pm 0.23$ & $0.764 \pm 0.041$ & $\mathbf{0.75 \pm 0.26}$ \\
            2D Soft2 & $0.846 \pm 0.094$ & $0.80 \pm 0.42$ & $\mathbf{0.828 \pm 0.064}$ & $0.74 \pm 0.31$ & $\mathbf{0.744 \pm 0.019}$ & $\mathbf{0.91 \pm 0.40}$ & $\mathbf{0.672 \pm 0.008}$ & $0.56 \pm 0.29$ & $\mathbf{0.772 \pm 0.041}$ & $0.75 \pm 0.33$ \\
            2D Soft3 & $0.845 \pm 0.092$ & $0.74 \pm 0.32$ & $0.822 \pm 0.067$ & $0.80 \pm 0.37$ & $0.724 \pm 0.038$ & $0.94 \pm 0.36$ & $0.658 \pm 0.015$ & $\mathbf{0.52 \pm 0.28}$ & $0.762 \pm 0.042$ & $0.75 \pm 0.30$ \\
            \bottomrule
        \end{tabular}%
    }
\end{table*}

Table~\ref{tab:soft_3D} shows results of the 3D model. The results indicate that 3D training did not benefit from soft labeling. The Hard Labels maintained the best Lesion GM HD95 (0.96 mm). All Soft Labels configurations worsen the results. This confirms that while soft labels refine 2D boundaries, they introduce counter-productive uncertainty into the 3D student's learning process.

\begin{table*}[t]
    \centering
    \caption{Soft Segmentation 3D Experiment -- Dice and HD95 3D Results. Best results are highlighted in bold.}
    \label{tab:soft_3D}
    \resizebox{\textwidth}{!}{%
        \begin{tabular}{lcccccccccc}
            \toprule
            & \multicolumn{2}{c}{\textbf{Healthy WM}} & \multicolumn{2}{c}{\textbf{Healthy GM}} & \multicolumn{2}{c}{\textbf{Lesion WM}} & \multicolumn{2}{c}{\textbf{Lesion GM}} & \multicolumn{2}{c}{\textbf{Total Average}} \\
            \cmidrule(lr){2-3} \cmidrule(lr){4-5} \cmidrule(lr){6-7} \cmidrule(lr){8-9} \cmidrule(lr){10-11}
            \textbf{Config} & \textbf{Dice} & \textbf{HD95} & \textbf{Dice} & \textbf{HD95} & \textbf{Dice} & \textbf{HD95} & \textbf{Dice} & \textbf{HD95} & \textbf{Dice} & \textbf{HD95} \\
            \midrule
            3D Hard & $\mathbf{0.838 \pm 0.106}$ & $\mathbf{0.72 \pm 0.14}$ & $\mathbf{0.827 \pm 0.091}$ & $0.93 \pm 0.21$ & $\mathbf{0.750 \pm 0.096}$ & $\mathbf{1.38 \pm 1.06}$ & $\mathbf{0.673 \pm 0.113}$ & $\mathbf{0.96 \pm 0.60}$ & $\mathbf{0.772 \pm 0.077}$ & $\mathbf{1.00 \pm 0.28}$ \\
            3D Soft1 & $0.833 \pm 0.108$ & $0.76 \pm 0.18$ & $0.818 \pm 0.091$ & $0.92 \pm 0.28$ & $0.728 \pm 0.090$ & $1.51 \pm 0.96$ & $0.650 \pm 0.085$ & $1.08 \pm 0.72$ & $0.757 \pm 0.085$ & $1.07 \pm 0.32$ \\
            3D Soft2 & $0.832 \pm 0.108$ & $0.79 \pm 0.19$ & $0.821 \pm 0.092$ & $\mathbf{0.92 \pm 0.23}$ & $0.733 \pm 0.093$ & $1.55 \pm 1.15$ & $0.666 \pm 0.090$ & $1.02 \pm 0.64$ & $0.763 \pm 0.078$ & $1.07 \pm 0.33$ \\
            3D Soft3 & $0.828 \pm 0.108$ & $0.77 \pm 0.09$ & $0.821 \pm 0.090$ & $0.99 \pm 0.22$ & $0.727 \pm 0.102$ & $1.53 \pm 1.01$ & $0.658 \pm 0.108$ & $0.97 \pm 0.56$ & $0.758 \pm 0.081$ & $1.07 \pm 0.32$ \\
            \bottomrule
        \end{tabular}%
    }
\end{table*}

\subsection{The Dimensionality Bridge (The 2D → 3D Framework)}

\subsubsection{The 2D Teacher Performance: The Slice-Wise Ceiling
}

Based on the extensive optimization, we established a Winning 2D Configuration comprising the AdamW optimizer, Dual-Channel input, Otsu background masking, Moderate Spatial Augmentation (Aug1), and Soft Segmentation 2. See Appendix~\ref{app:otsu} for the Otsu Masking experiment, Appendix~\ref{app:optimizer} for the optimizer experiment and Appendix~\ref{app:synergy} for the synergy analysis. Additionaly, Appendix~\ref{app:strategy} explains how we combined ablation study and exploratory analysis to obtain rigorous results while limiting computational time. To maximize the quality of the pseudo-labels generated for the student model, we further enhanced this configuration using the nnU-Net Test Time Augmentation (TTA) as shown in Table~\ref{tab:win_2D}.

\begin{table*}[t]
    \centering
    \caption{Winning 2D Combination -- Dice and HD95 2D Results on the Validation Set. Best results are highlighted in bold.}
    \label{tab:win_2D}
    \resizebox{\textwidth}{!}{%
        \begin{tabular}{lcccccccccc}
            \toprule
            & \multicolumn{2}{c}{\textbf{Healthy WM}} & \multicolumn{2}{c}{\textbf{Healthy GM}} & \multicolumn{2}{c}{\textbf{Lesion WM}} & \multicolumn{2}{c}{\textbf{Lesion GM}} & \multicolumn{2}{c}{\textbf{Total Average}} \\
            \cmidrule(lr){2-3} \cmidrule(lr){4-5} \cmidrule(lr){6-7} \cmidrule(lr){8-9} \cmidrule(lr){10-11}
            \textbf{Config} & \textbf{Dice} & \textbf{HD95} & \textbf{Dice} & \textbf{HD95} & \textbf{Dice} & \textbf{HD95} & \textbf{Dice} & \textbf{HD95} & \textbf{Dice} & \textbf{HD95} \\
            \midrule
            Winner & $0.846 \pm 0.094$ & $0.80 \pm 0.42$ & $0.828 \pm 0.064$ & $0.74 \pm 0.31$ & $\mathbf{0.744 \pm 0.019}$ & $0.91 \pm 0.40$ & $0.672 \pm 0.008$ & $0.56 \pm 0.29$ & $0.772 \pm 0.041$ & $0.75 \pm 0.33$ \\
            Winner + TTA & $\mathbf{0.848 \pm 0.090}$ & $\mathbf{0.79 \pm 0.41}$ & $\mathbf{0.829 \pm 0.062}$ & $\mathbf{0.72 \pm 0.30}$ & $0.740 \pm 0.040$ & $\mathbf{0.83 \pm 0.31}$ & $\mathbf{0.678 \pm 0.006}$ & $\mathbf{0.53 \pm 0.26}$ & $\mathbf{0.774 \pm 0.036}$ & $\mathbf{0.72 \pm 0.29}$ \\
            \bottomrule
        \end{tabular}%
    }
\end{table*}

This optimized 2D Teacher achieved the highest slice-wise accuracy in our study, serving as the slice-wise performance ceiling for our 2D models. As detailed in Table~\ref{tab:win_2D}, the model reached a combined average Dice of 0.774, with notable stability in the challenging Lesion GM class (0.678±0.006). It showed steady performances on the test set with an average Dice of 0.752 and even better performance when using TTA (Test-Time Augmentation). Those results are confirmed by the performance obtained on the held out test set, as shown in Table~\ref{tab:win_2D_test}. Furthermore, ensembling the four fold-specific models during inference provided an additional performance gain, improving overall results by nearly 2 percentage points.

\begin{table*}[t]
    \centering
    \caption{Winning 2D Combination -- Dice and HD95 2D Results on the Test Set. Best results are highlighted in bold.}
    \label{tab:win_2D_test}
    \resizebox{\textwidth}{!}{%
        \begin{tabular}{lcccccccccc}
            \toprule
            & \multicolumn{2}{c}{\textbf{Healthy WM}} & \multicolumn{2}{c}{\textbf{Healthy GM}} & \multicolumn{2}{c}{\textbf{Lesion WM}} & \multicolumn{2}{c}{\textbf{Lesion GM}} & \multicolumn{2}{c}{\textbf{Total Average}} \\
            \cmidrule(lr){2-3} \cmidrule(lr){4-5} \cmidrule(lr){6-7} \cmidrule(lr){8-9} \cmidrule(lr){10-11}
            \textbf{Config} & \textbf{Dice} & \textbf{HD95} & \textbf{Dice} & \textbf{HD95} & \textbf{Dice} & \textbf{HD95} & \textbf{Dice} & \textbf{HD95} & \textbf{Dice} & \textbf{HD95} \\
            \midrule
            Winner + TTA (Test) & $0.876 \pm 0.006$ & $\mathbf{0.71 \pm 0.06}$ & $0.768 \pm 0.008$ & $0.93 \pm 0.06$ & $0.706 \pm 0.018$ & $1.11 \pm 0.10$ & $0.658 \pm 0.020$ & $0.60 \pm 0.02$ & $0.752 \pm 0.007$ & $0.84 \pm 0.03$ \\
            Winner + TTA + Ensemble (Test) & $\mathbf{0.889}$ & $0.71$ & $\mathbf{0.779}$ & $\mathbf{0.85}$ & $\mathbf{0.735}$ & $\mathbf{0.85}$ & $\mathbf{0.674}$ & $\mathbf{0.53}$ & $\mathbf{0.769}$ & $\mathbf{0.74}$ \\
            \bottomrule
        \end{tabular}%
    }
\end{table*}

\begin{figure}
    \centering
    \includegraphics[width=1\linewidth]{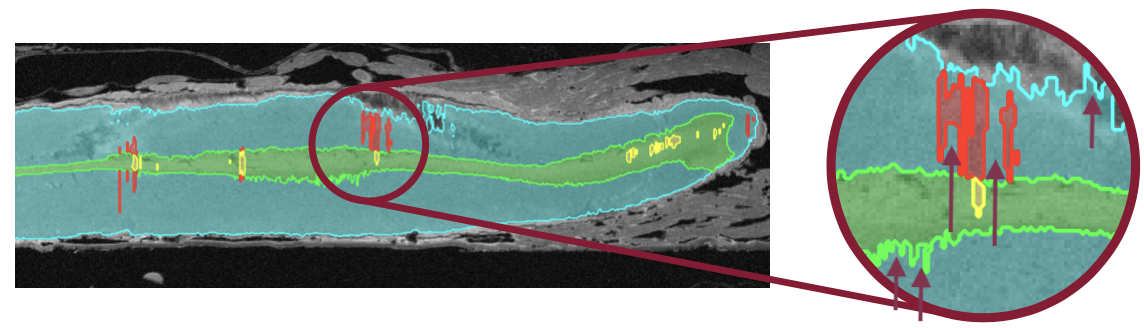}
    \caption{Illustration of the jitter noise on a sagittal plane (Cyan: Healthy WM, Green: Healthy GM, Red: Lesion WM, Yellow: Lesion GM).}
    \label{fig:noise}
\end{figure}

However, despite this high quantitative performance on individual slices, when reconstructed into volumes the 2D predictions exhibited characteristic inconsistencies on the stack axis. Because the 2D network treats each axial slice as an independent sample, it lacks the longitudinal context to ensure anatomical continuity along the Z-axis. The resulting predictions are jagged and discontinuous (Figure~\ref{fig:noise}). Although these 2D cross-sections are accurate, the 3D spatial predictions are biologically implausible, which underscores the necessity of a trained 3D model.

\subsubsection{The 3D Student: Volumetric Refinement}

Regarding the 3D stream, since no additive experiments yielded improvements, the the Winning 3D Configuration is a streamlined Base architecture (AdamW + Dual-Channel) utilizing Patch size 5 (192×208×64), prioritizing axial completeness over longitudinal extent. An ablation study justifying this geometric trade-off is provided in Appendix~\ref{app:patch_size}.

\begin{table*}[t]
    \centering
    \caption{Winning 3D Combination -- Dice and HD95 3D Results on the Validation Set. Best results are highlighted in bold.}
    \label{tab:win_3D}
    \resizebox{\textwidth}{!}{%
        \begin{tabular}{lcccccccccc}
            \toprule
            & \multicolumn{2}{c}{\textbf{Healthy WM}} & \multicolumn{2}{c}{\textbf{Healthy GM}} & \multicolumn{2}{c}{\textbf{Lesion WM}} & \multicolumn{2}{c}{\textbf{Lesion GM}} & \multicolumn{2}{c}{\textbf{Total Average}} \\
            \cmidrule(lr){2-3} \cmidrule(lr){4-5} \cmidrule(lr){6-7} \cmidrule(lr){8-9} \cmidrule(lr){10-11}
            \textbf{Config} & \textbf{Dice} & \textbf{HD95} & \textbf{Dice} & \textbf{HD95} & \textbf{Dice} & \textbf{HD95} & \textbf{Dice} & \textbf{HD95} & \textbf{Dice} & \textbf{HD95} \\
            \midrule
            3D Winner & $0.838 \pm 0.092$ & $0.72 \pm 0.13$ & $\mathbf{0.827 \pm 0.078}$ & $\mathbf{0.93 \pm 0.18}$ & $0.750 \pm 0.083$ & $1.38 \pm 0.92$ & $\mathbf{0.673 \pm 0.098}$ & $\mathbf{0.96 \pm 0.52}$ & $0.772 \pm 0.079$ & $1.00 \pm 0.35$ \\
            3D Winner + TTA & $\mathbf{0.838 \pm 0.092}$ & $\mathbf{0.72 \pm 0.13}$ & $0.827 \pm 0.078$ & $\mathbf{0.93 \pm 0.18}$ & $\mathbf{0.750 \pm 0.083}$ & $\mathbf{1.38 \pm 0.92}$ & $0.673 \pm 0.098$ & $\mathbf{0.96 \pm 0.52}$ & $\mathbf{0.772 \pm 0.079}$ & $\mathbf{1.00 \pm 0.35}$ \\
            \bottomrule
        \end{tabular}%
    }
\end{table*}

Table~\ref{tab:win_3D} shows the 3D Student achieved a total average Dice of 0.772 with TTA, which is slightly less than the 0.774 score of the 2D Teacher Model. Interestingly, we can note that the TTA did not significantly change the results of the 3D models. The transition to 3D resulted in a slight increase in the HD95 boundary distance, shifting from 0.72 mm in the 2D Teacher to 1.00 mm in the 3D Student. These results are confirmed by the performance obtained on the held-out test set, as shown in Table~\ref{tab:win_3D_test}. Finally, on the test set, the use of all 4-fold models to run ensemble inference improved the results yet again from an average Dice Score of 0.776 to 0.783.

\begin{figure}[htbp]
    \centering
    \begin{tabular}{ccc}
        (a) & (b) & (c) \\
        \includegraphics[height=0.09\textheight, angle=90, origin=c]{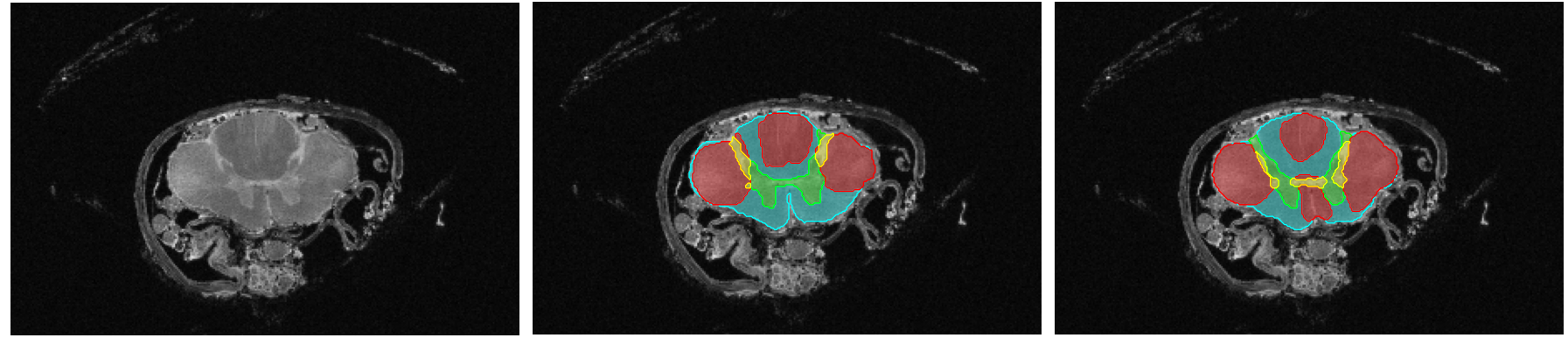} &
        \includegraphics[height=0.09\textheight, angle=90, origin=c]{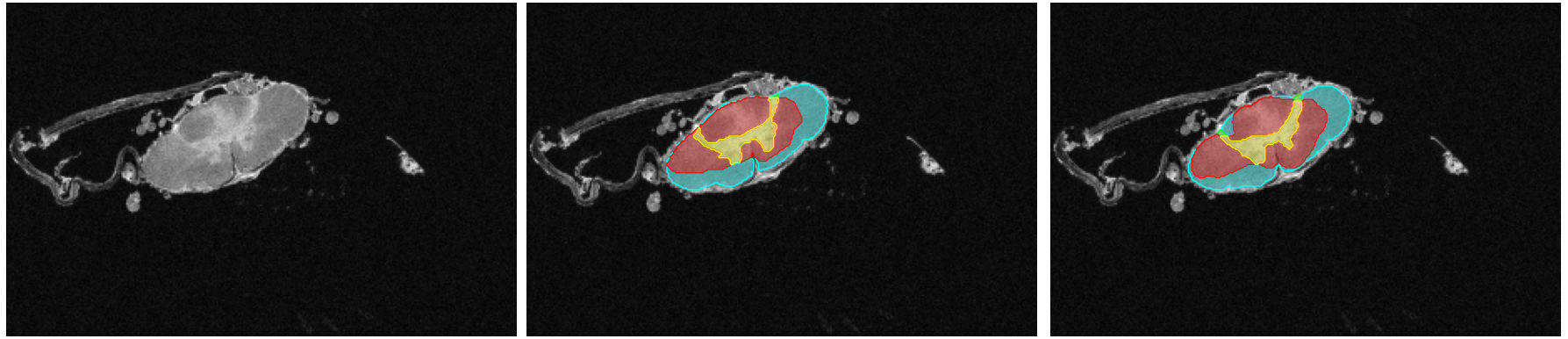} &
        \includegraphics[height=0.09\textheight, angle=90, origin=c]{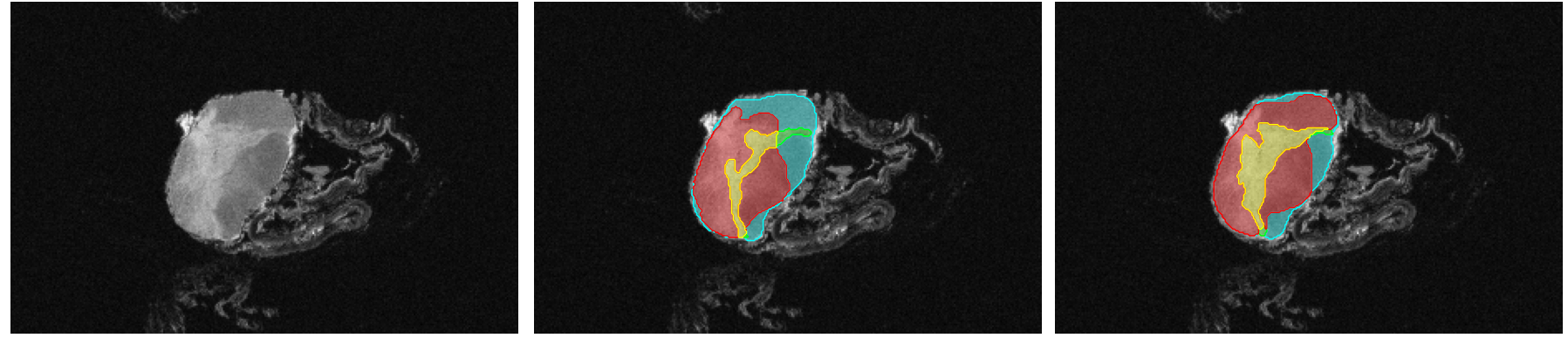} \\
    \end{tabular}
    \vspace{-0.5em} 
    \caption{Segmentation results on the test set. Each panel shows (from bottom to top) the magnitude, the ground truth, and the prediction. (Cyan: Healthy WM, Green: Healthy GM, Red: Lesion WM, Yellow: Lesion GM)}
    \label{fig:test_results_combined}
\end{figure}

To complement these quantitative metrics, Figure~\ref{fig:test_results_combined} provides a qualitative visual assessment of the model's predictions on the test set. While the model generally performs well, several common failure modes can be observed. Panel (a) illustrates a case of oversegmentation, where the model incorrectly predicts an additional lesion in the anterior part of the cord. Conversely, panel (b) highlights an instance of undersegmentation, where the model incorrectly deems a portion of the lesioned white matter in the posterior part as healthy tissue. Furthermore, panel (c) demonstrates a particularly challenging scenario where the model struggles significantly with gray matter boundaries that are rendered completely invisible by a lesion on the magnitude image; as a result, it also overestimates the size of the white matter lesion. This difficulty in delineating ambiguous boundaries in the presence of severe signal alterations reflects a general behavior observed across the model's predictions.

\begin{table*}[t]
    \centering
    \caption{Winning 3D Combination -- Dice and HD95 3D Results on the Test Set. Best results are highlighted in bold.}
    \label{tab:win_3D_test}
    \resizebox{\textwidth}{!}{%
        \begin{tabular}{lcccccccccc}
            \toprule
            & \multicolumn{2}{c}{\textbf{Healthy WM}} & \multicolumn{2}{c}{\textbf{Healthy GM}} & \multicolumn{2}{c}{\textbf{Lesion WM}} & \multicolumn{2}{c}{\textbf{Lesion GM}} & \multicolumn{2}{c}{\textbf{Total Average}} \\
            \cmidrule(lr){2-3} \cmidrule(lr){4-5} \cmidrule(lr){6-7} \cmidrule(lr){8-9} \cmidrule(lr){10-11}
            \textbf{Config} & \textbf{Dice} & \textbf{HD95} & \textbf{Dice} & \textbf{HD95} & \textbf{Dice} & \textbf{HD95} & \textbf{Dice} & \textbf{HD95} & \textbf{Dice} & \textbf{HD95} \\
            \midrule
            3D Winner + TTA (Test) & $0.888 \pm 0.003$ & $0.63 \pm 0.11$ & $0.796 \pm 0.010$ & $\mathbf{0.67 \pm 0.19}$ & $0.741 \pm 0.007$ & $1.11 \pm 0.13$ & $0.680 \pm 0.007$ & $0.58 \pm 0.04$ & $0.776 \pm 0.006$ & $0.75 \pm 0.09$ \\
            3D Winner + TTA + Ensemble (Test) & $\mathbf{0.891}$ & $\mathbf{0.53}$ & $\mathbf{0.800}$ & $0.84$ & $\mathbf{0.751}$ & $\mathbf{1.03}$ & $\mathbf{0.690}$ & $\mathbf{0.57}$ & $\mathbf{0.783}$ & $\mathbf{0.74}$ \\
            \bottomrule
        \end{tabular}%
    }
\end{table*}

\subsubsection{Analysis of Volumetric Consistency}

The transition to 3D provided a critical gain in volumetric plausibility that slice-based metrics fail to fully capture. The 2D model's predictions are characterized by high-frequency jitter along the Z-axis.

\begin{figure}[htbp] 
    \centering
    
    % First Subfigure (Left)
    \begin{subfigure}[t]{0.45\linewidth} 
        \centering
        % We force the height to a specific size (e.g., 4cm)
        \includegraphics[height=11.5cm, keepaspectratio]{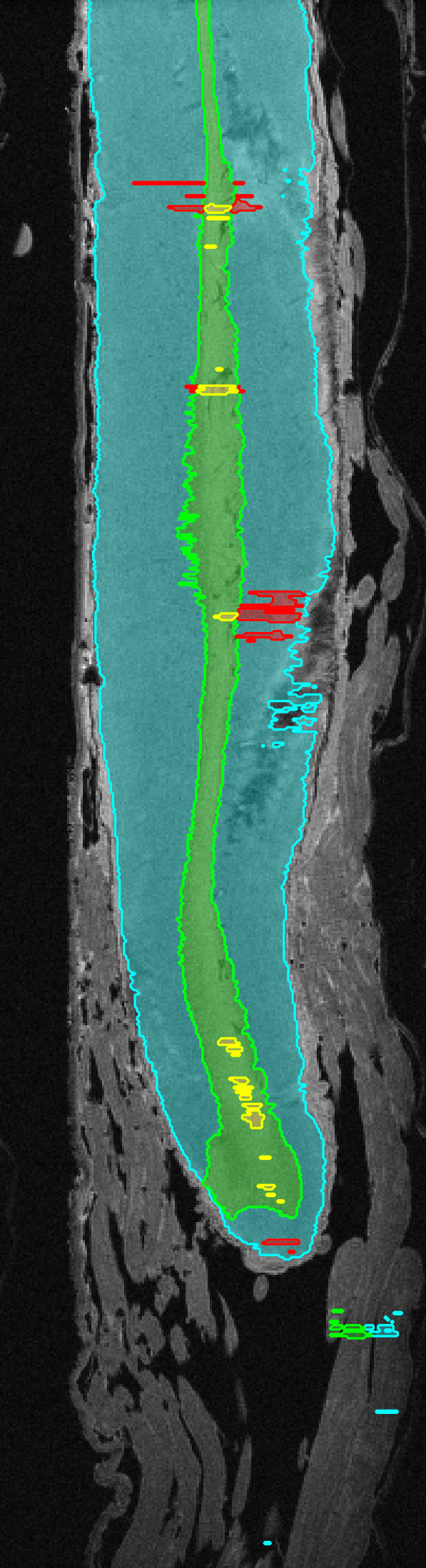}
        \caption{Pseudo GT made up of stacked 2D slices} 
        \label{fig:sagittal_noise}
    \end{subfigure}% <-- This % symbol prevents an invisible space!
    \hspace{0.5cm} % <-- This gives a fixed 0.5cm gap instead of a massive \hfill gap
    % Second Subfigure (Right)
    \begin{subfigure}[t]{0.45\linewidth}
        \centering
        % We use the exact same height here!
        \includegraphics[height=11.5cm, keepaspectratio]{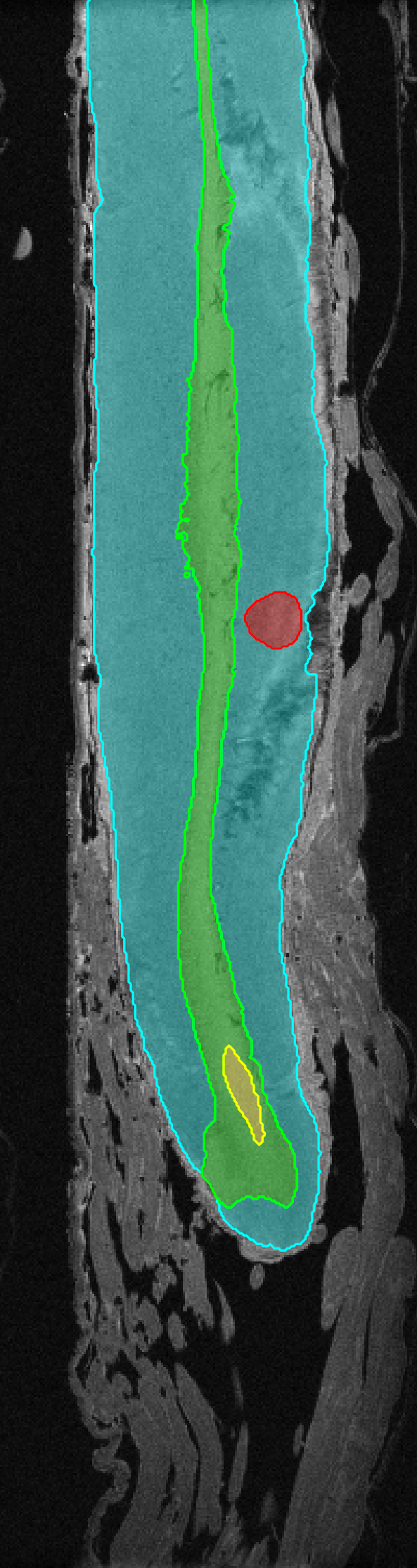}
        \caption{3D Prediction}
        \label{fig:sagittal_smooth}
    \end{subfigure}
    
    \caption{Sagittal view of a rough Pseudo Ground Truth and a smooth Prediction (Cyan: Healthy WM, Green: Healthy GM, Red: Lesion WM, Yellow: Lesion GM)}
    \label{fig:gt_vs_3D_pred}
\end{figure}

Geometric Continuity ($DSC_{z}$): The naive stacking of 2D predictions yielded a low inter-slice Dice Score ($DSC_{z}$) of 0.785 (higher is smoother), quantitatively reflecting significant discontinuity between adjacent slices. In contrast, the 3D Student achieved a $DSC_{z}$ of 0.900 (without TTA and Ensembling), confirming that the model successfully learned the continuous tubular anatomy of the spinal cord rather than treating it as a disjointed stack of slices. This is illustrated qualitatively by Figure~\ref{fig:gt_vs_3D_pred}. 

Boundary Precision (HD95): The cost of this increased volumetric smoothness is quantitatively captured by the surface distance metrics. As shown in Table~\ref{tab:win_3D}, the 3D Student recorded a higher distance error compared to the highly optimized 2D Teacher (3D: 1.00 mm vs. 2D: 0.72 mm, on the validation set with TTA). Rather than indicating a failure of the 3D model, this reveals a fundamental trade-off in the Sparse-to-Dense pipeline. The 2D Teacher, unburdened by longitudinal constraints, aggressively over-optimizes for exact in-plane contour adherence. In contrast, the 3D Student sacrifices a fraction of a millimeter of strict 2D slice adherence in order to enforce biological plausibility. It filters out the jagged, noisy edges typical of slice-independent predictions to generate boundaries that are compliant with the continuous, tubular 3D anatomical surface.

\section{Discussion}

In this work, we presented a 2D Sparse to 3D Dense framework to bridge the annotation bottleneck in high-resolution medical image volume – in this case, ex vivo spinal cord MRI with multiple segmentation classes. While the transition from sparse 2D annotations to dense 3D volumes is a known strategy, our ablation study revealed counter-intuitive findings regarding training dynamics. We demonstrated that techniques widely accepted in computer vision (and even in medical imaging literature) do not universally translate to this specific domain in the case of high resolution spinal cord segmentation. Most notably, we observed a distinct divergence between the optimization needs of the 2D Teacher and the 3D Student, suggesting that the dimensionality alters the model's susceptibility to noise and shortcut learning, making additional regularization techniques possibly harmful.

\subsection{Human Perception vs. Machine Statistics}

A critical finding of our study is the danger of preprocessing techniques designed for human visual clarity. We believed that Contrast Limited Adaptive Histogram Equalization (CLAHE) and contrast stretching (methods shown to improve segmentation in other MRI domains by \cite{Yoshimi2024-gj}) would aid the network in resolving subtle gray matter boundaries. However, our results showed the opposite: these techniques can cause significant performance degradation.

We attribute this to a fundamental disconnect between human and machine vision. Human raters rely on local image contrast enhancement to distinguish boundaries within a wide dynamic range. In contrast, the nnU-Net architecture relies on consistent global intensity statistics (mean and standard deviation) to characterize tissue types. By artificially manipulating the histogram to maximize local contrast, CLAHE effectively eliminates these global statistical cues, presenting the network with out-of-distribution data. This serves as a cautionary tale: preprocessing that optimizes radiological interpretability can be actively detrimental to computational segmentation pipelines.

\subsection{Multi-Modal Synergy and Attention Mechanisms}

Our experiments confirm that the integration of multi-modal data is not merely additive but synergistic, effectively helping to resolve the ``detection-localization'' trade-off in spinal cord segmentation. While T2*-weighted magnitude imaging provides the hyperintense signal necessary for lesion detection, it frequently lacks the contrast required to situate those lesions within the correct tissue type. The inclusion of phase data provided the necessary orthogonal information for anatomical localization, particularly for resolving the ambiguous gray matter/white matter boundaries of lesions. This aligns with the biophysical consensus established by \cite{Duyn2007-fi}, who demonstrated that phase imaging can yield a Contrast-to-Noise Ratio (CNR) between gray matter and white matter up to ten times higher than magnitude sequences. Our results mirror findings in similar deep learning applications, such as RimNet \citep{Gros2021-gx}, confirming that fusing phase structure with magnitude intensity is critical for stabilizing predictions in complex multi-class tasks. A biophysical explanation is provided in Appendix~\ref{app:biophysics}. 

However, the method of constraining this input space revealed a critical divergence between 2D and 3D training dynamics. We applied 'hard attention' via Otsu background masking to focus the models on the spinal cord. In the 2D Teacher, this masking was beneficial, effectively solving a class imbalance problem by eliminating false positives in the background environment. Conversely, this same technique degraded performance in the 3D Student (see Appendix~\ref{app:otsu}). We hypothesize that the 3D model, by leveraging volumetric convolution kernels, possesses an inherent robustness to scattered background noise that 2D models lack. The hard masking likely removed subtle noise patterns or global spatial cues that the 3D model utilizes for volumetric alignment. This suggests that while 2D models benefit from rigid, pre-calculated constraints, 3D networks thrive on unmasked contexts, favoring soft, learned attention over manual feature engineering.

\subsection{The Regularization: 2D-to-3D Bridging}
Perhaps the most relevant insight from our work is the behavior of regularization techniques across the dimension gap. We observed that variations which significantly boosted the 2D Teacher (specifically strong spatial augmentation and soft labeling) failed to improve, or even degraded, the 3D Student.

This particular behavior contradicts standard literature in both general computer vision \citep{Geirhos2020-an, Geirhos2018-cc} and in the medical field \citep{Chaitanya2020-ae, Oza2022-ba}, which suggests that geometric augmentation and boundary uncertainty modeling are beneficial for 3D medical segmentation. We propose that in a Sparse-to-Dense framework, the bridging process can make regularization techniques harmful.

\begin{itemize}
    \item Shortcut Learning: The 2D model, seeing only sparse slices, is highly prone to shortcut learning (e.g., memorizing that the spinal cord is always in the center). Aggressive spatial augmentation was required to break these spurious correlations. The 3D Student, however, is trained on dense pseudo-labels generated by an ensemble of 2D models. The 3D model is confronted with a vast and dense amount of data. This makes it less likely that it relies on shortcuts. 
    \item Uncertainty \& Over-Smoothing: Similarly, while Soft Labeling helped the 2D model handle subjective lesion boundaries, the 3D Student did not benefit from it. The pseudo-labels used for 3D training are already the average of multiple models (via cross-validation ensembling), which inherently smooths out uncertainty. Applying additional soft-loss smoothing on top of this likely leads to over-regularization, eroding the precision of the segmentation.
\end{itemize}

\subsection{Limitations}

While our framework successfully leverages sparse annotations to train robust volumetric models, several limitations warrant discussion. First, the evaluation of the 3D Student model is inherently constrained by the absence of a dense volumetric ground truth. We rely on sparse axial slices for quantitative validation; thus, our metrics (Dice Score, HD95) serve as proxies for volumetric performance.

Second, this study utilizes high-resolution ex vivo MRI data acquired at 9.4T. While this provides superior anatomical detail compared to clinical scans, ex vivo tissue lacks the physiological noise (motion artifacts, blood flow pulsatility) and partial volume effects typical of \textit{in vivo} acquisitions. Consequently, the specific preprocessing interactions and performance baselines reported here may not translate directly to lower-resolution \textit{in vivo} clinical datasets without further adaptation.

Third, the Student model is fundamentally bound by the quality of the pseudo-labels generated by the Teacher. Although Ensembling and Test Time Augmentation were employed to maximize label fidelity, any systematic bias or hallucination present in the 2D Teacher’s output (particularly in regions of poor contrast) is likely propagated to the 3D Student.

\section{Conclusion}

We presented a scalable pipeline that effectively bridges the gap between sparse 2D manual inputs and robust 3D volumetric segmentation. Our extensive ablation and additive studies challenge some of the best and intuitive practices of computer vision techniques in medical imaging. We demonstrated that preprocessing based on human perception can be harmful, and that the optimal training strategy changes fundamentally when moving from a 2D Teacher to a 3D Student. The transition to 3D does not merely add a dimension; it fundamentally alters the noise profile and regularization needs of the network. Our findings indicate that the strong regularization strategies required to train data-efficient 2D models become detrimental when applied to 3D architectures trained on dense pseudo-labels. Future work in weakly supervised segmentation should view the 2D-to-3D transition not just as an inference trick, but as a shift in learning dynamics that requires a distinct, conservative hyperparameter strategy rather than the blind transfer of 2D augmentation techniques.

\acks{We thank Maxime Donadieu for his assistance with postmortem sample preparation and the optimization of MRI sequences on the 9.4T scanner. Funded by the SensoriMotor Rehabilitation Research Team (SMRRT) of the Canadian Institute of Health Research, the National MS Society [FG1892A1/1], the Fonds de Recherche du Québec - Santé (FRQS), the Quebec BioImaging Network (QBIN), the Natural Sciences and Engineering Research Council of Canada (NSERC) and the Swiss National Science Foundation and National Multiple Sclerosis Society. This research was supported in part by the Intramural Research Program of the National Institutes of Health (NIH), including the NINDS Intramural Research Program. The contributions of the NIH authors are considered Works of the United States Government. The findings and conclusions presented in this paper are those of the authors and do not necessarily reflect the views of the NIH or the U.S. Department of Health and Human Services.}

% Ethical Standards.
% Please edit with the appropriate ethics considerations for your work. Include any pertinent IRB information, etc.
%
% Please note that the submission requirements included:
% The work presented must follow appropriate ethical standards in conducting research and writing the manuscript, following all applicable laws and regulations regarding treatment of animals or human subjects.
\ethics{The work follows appropriate ethical standards in conducting research and writing the manuscript, following all applicable laws and regulations regarding treatment of human subjects.}

% Conflict of Interest
% Declaration of possible conflicts of interest: Authors must disclose any financial, organisational, commercial or personal conflicts of interest that might bias their work.
% If no conflicts, please say ``We declare we don't have conflicts of interest.''
\coi{Daniel S. Reich receives research funding from Sanofi. Irene Cortese owns shares in Nouscom AG and Keires AG (outside the scope of this work). All other authors declare no conflicts of interest.}

% Data availability
\data{The code and models are available at ~\url{https://github.com/ivadomed/model_seg_sc-gm-lesion_human_ms_exvivo_t2star}. The dataset supporting the findings of this study will be made publicly available upon acceptance of this publication and hosted on Borealis, the Canadian Dataverse Repository.}

\bibliography{references}

% needed in general before appenDices.
% \newpage

% Appendix is optional
\clearpage
\appendix
\renewcommand{\theHsection}{A.\thesection}

\section{Experimental Strategy: Hybrid Ablation and Exploration}
\label{app:strategy}
Unlike standard additive studies that start from a ``Vanilla'' default and incrementally add features, we established a Strong Baseline configuration derived from preliminary screening. This baseline represents a model performing better than the default nnU-Net, incorporating specific input modalities and preprocessing steps. To systematically validate this configuration and explore complex hyperparameters, we adopted a hybrid experimental strategy:

\subsubsection*{Subtractive Ablation (Validation of Core Components)} 
To demonstrate the contribution of specific binary design choices (e.g., the inclusion of phase data or Otsu masking) and Configuration elements that passed the screening, we employed a subtractive approach on a Baseline configuration. We compare the Strong Baseline against a version of itself with a single component removed. A drop in performance validates the necessity of that component. 
\begin{itemize}
    \item For the 2D models, the Baseline was: AdamW, Magnitude+Phase modalities, Otsu Masking, No Additional Preprocessing, Spatial Augmentation Strategy 1 and Hard Labels. 
    \item For the 3D models, the Baseline was: AdamW, Magnitude+Phase modalities, No Preprocessing (other than nnU-Net additional preprocessing), No Spatial Augmentation, Hard Labels and Patch Size 5. 
\end{itemize}

\subsubsection*{Additive \& Parametric Exploration (Complex Hyperparameters)} 
For components with a complex search space, such as the intensity of spatial augmentation or the temperature of soft-label losses, a subtractive approach is insufficient. For these, we performed additive or parametric studies, exploring how varying specific hyperparameters impacts the model behavior relative to the baseline. This additive approach also includes the training configuration elements that did not pass the first screening. 
\begin{itemize}
    \item For the 2D models, the additive experiments are: CLAHE and Gamma Stretching, Spatial Augmentations and Soft Labels 
    \item For the 3D models, the additive experiments are: CLAHE and Gamma Stretching, Otsu Masking, Spatial Augmentations and Soft Labels.
\end{itemize}

\begin{table*}[t]
    \centering
    \caption{Patch Size Experiment -- Dice and HG95 3D Results. Best results are highlighted in bold.}
    \label{tab:patch_3D}
    \resizebox{\textwidth}{!}{%
        \begin{tabular}{lcccccccccc}
            \toprule
            & \multicolumn{2}{c}{\textbf{WM}} & \multicolumn{2}{c}{\textbf{GM}} & \multicolumn{2}{c}{\textbf{Lesion WM}} & \multicolumn{2}{c}{\textbf{Lesion GM}} & \multicolumn{2}{c}{\textbf{Total Average}} \\
            \cmidrule(lr){2-3} \cmidrule(lr){4-5} \cmidrule(lr){6-7} \cmidrule(lr){8-9} \cmidrule(lr){10-11}
            \textbf{Config} & \textbf{Dice} & \textbf{HG95} & \textbf{Dice} & \textbf{HG95} & \textbf{Dice} & \textbf{HG95} & \textbf{Dice} & \textbf{HG95} & \textbf{Dice} & \textbf{HG95} \\
            \midrule
            Patch size 1 & $0.819 \pm 0.116$ & $0.83 \pm 0.16$ & $0.811 \pm 0.101$ & $0.76 \pm 0.31$ & $0.716 \pm 0.109$ & $1.38 \pm 0.66$ & $0.654 \pm 0.115$ & $0.87 \pm 0.58$ & $0.750 \pm 0.110$ & $0.96 \pm 0.43$ \\
            Patch size 2 & $0.832 \pm 0.107$ & $0.75 \pm 0.09$ & $0.822 \pm 0.092$ & $0.77 \pm 0.37$ & $0.729 \pm 0.103$ & $1.22 \pm 0.76$ & $0.657 \pm 0.115$ & $0.74 \pm 0.46$ & $0.760 \pm 0.104$ & $0.87 \pm 0.42$ \\
            Patch size 3 & $0.833 \pm 0.107$ & $0.73 \pm 0.16$ & $0.823 \pm 0.087$ & $0.78 \pm 0.27$ & $0.728 \pm 0.101$ & $1.19 \pm 0.67$ & $0.657 \pm 0.109$ & $0.76 \pm 0.51$ & $0.760 \pm 0.101$ & $0.87 \pm 0.40$ \\
            Patch size 4 & $0.830 \pm 0.106$ & $\mathbf{0.72 \pm 0.08}$ & $0.825 \pm 0.090$ & $\mathbf{0.75 \pm 0.27}$ & $0.727 \pm 0.095$ & $\mathbf{1.12 \pm 0.69}$ & $0.664 \pm 0.103$ & $0.71 \pm 0.48$ & $0.762 \pm 0.098$ & $\mathbf{0.82 \pm 0.38}$ \\
            Patch size 5 & $\mathbf{0.834 \pm 0.106}$ & $0.75 \pm 0.08$ & $\mathbf{0.826 \pm 0.090}$ & $0.77 \pm 0.32$ & $\mathbf{0.733 \pm 0.088}$ & $1.14 \pm 0.68$ & $\mathbf{0.667 \pm 0.100}$ & $\mathbf{0.71 \pm 0.45}$ & $\mathbf{0.765 \pm 0.096}$ & $0.84 \pm 0.38$ \\
            \bottomrule
        \end{tabular}%
    }
\end{table*}

\section{3D Patch Size Optimization and Geometric Constraints}
\label{app:patch_size}

For the 2D Teacher, the computational efficiency of slice-wise processing allowed us to utilize a patch size covering the entire image dimensions, effectively eliminating the need for hyperparameter tuning in this regard. However, the transition to 3D volumetric training introduces significant memory constraints, necessitating the definition of a specific input patch size that dictates the network's Field of View. This choice imposes a critical geometric trade-off specific to the spinal cord's tubular anatomy: balancing the need for complete axial context (X-Y planes) against the benefits of extensive longitudinal continuity (Z-axis).

\subsubsection*{Motivation} 

In 3D volumetric segmentation, the patch size dictates the Field of View the network sees during training. This introduces a critical trade-off: larger patches provide more anatomical context but increase computational memory requirements. For the spinal cord, which is characterized by a ``tubular'' geometry (long and continuous in the Z-axis, but small and centered in the X-Y axial plane), determining the optimal balance between axial context and longitudinal continuity is essential. We needed to determine if the model benefited more from seeing a long segment of the cord (High Z-dimension) or the complete cross-sectional anatomy (High X-Y dimension).

\subsubsection*{Methodology} 
We evaluated five distinct patch size configurations to identify the optimal input dimensions for the 3D Student model. These configurations, detailed in Table~\ref{tab:patch_3D}, ranged from ``pencil-like'' patches (Patch Size 1: maximizing Z-depth at the cost of axial width) to ``slab-like'' patches (Patch Size 5: maximizing axial width to cover the full spine cross-section, with moderate Z-depth). All models were trained on the dense pseudo-labels generated by the best 2D ensemble.

\subsubsection*{Results}
Table~\ref{tab:patch_3D} shows the results of the 3D models. We observed that prioritizing the longitudinal Z-dimension at the expense of axial context was detrimental to performance. ``Patch Size 1'' (which had the largest Z-dimension of 144 but the smallest axial dimensions) yielded the lowest global Dice Score of 0.750. Conversely, ``Patch Size 5,'' which prioritized a large axial field of view (192×208) with a reduced Z-dimension (64), achieved the highest performance with a total average Dice of 0.765. This configuration outperformed intermediate balances (Patch Sizes 2, 3, and 4) and achieved the highest stability for the challenging Lesion WM class (0.733).

\begin{figure}[ht!]
    \centering
    
    % --- Row 1 ---
    \begin{subfigure}[t]{0.4\columnwidth} % <--- Changed [b] to [t]
        \centering
        \includegraphics[width=\textwidth]{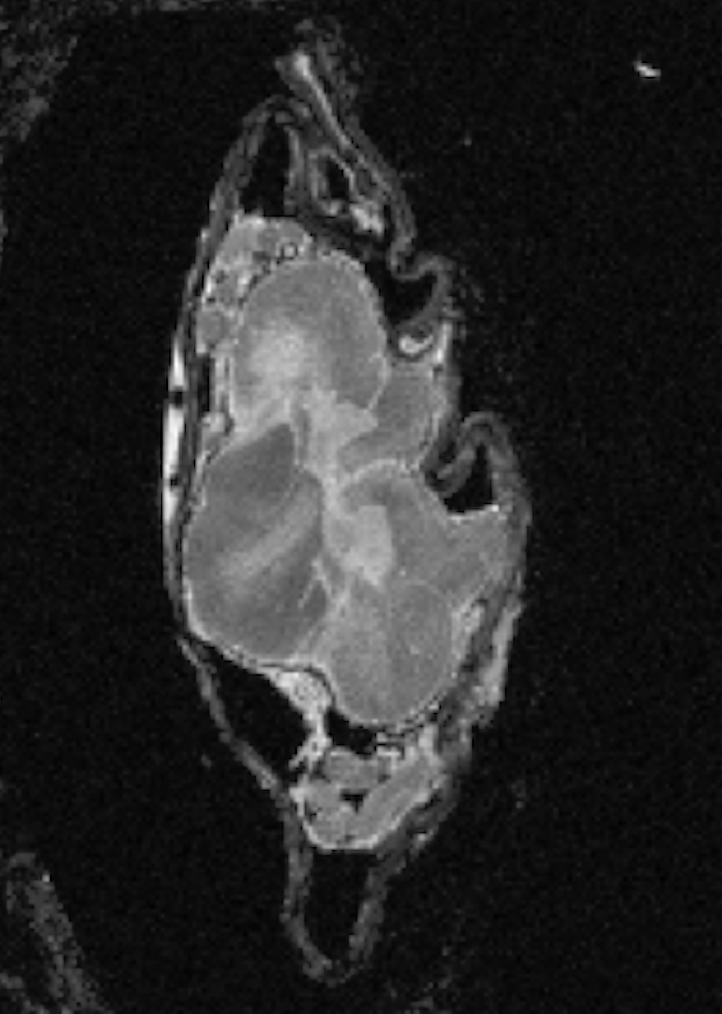} 
        \caption{Original Magnitude}
        \label{fig:otsu_mag_orig}
    \end{subfigure}%
    \hspace{0.5cm}
    \begin{subfigure}[t]{0.4\columnwidth} % <--- Changed [b] to [t]
        \centering
        \includegraphics[width=\textwidth]{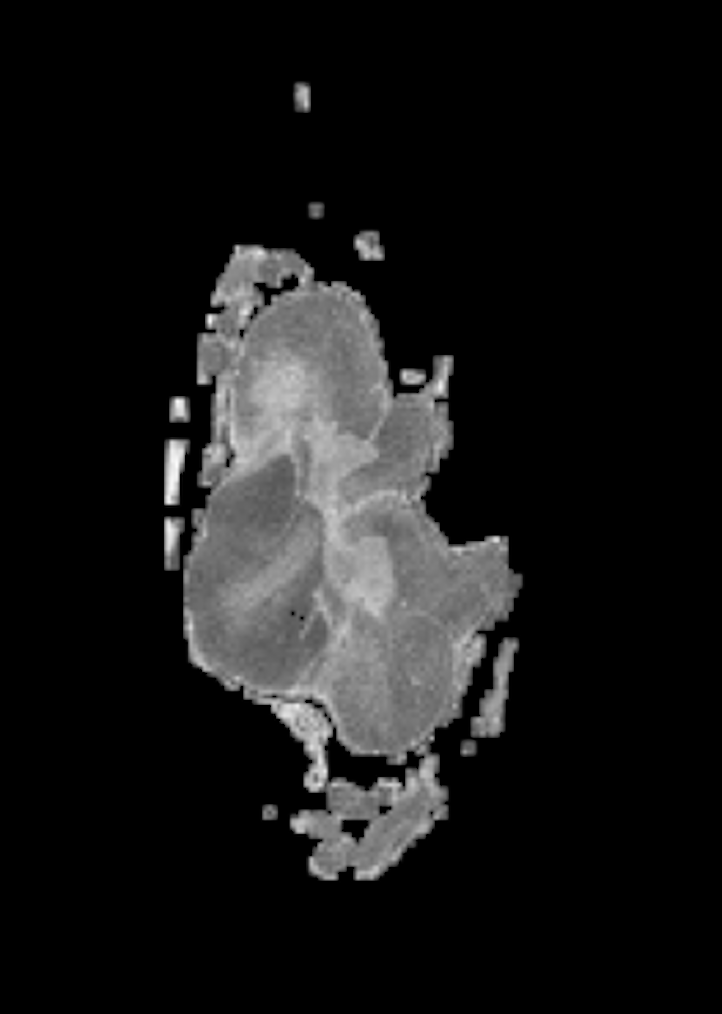}
        \caption{Otsu Masked Magnitude}
        \label{fig:otsu_mag}
    \end{subfigure}
    
    \vspace{0.5cm} 
    
    % --- Row 2 ---
    \begin{subfigure}[t]{0.4\columnwidth} % <--- Changed [b] to [t]
        \centering
        \includegraphics[width=\textwidth]{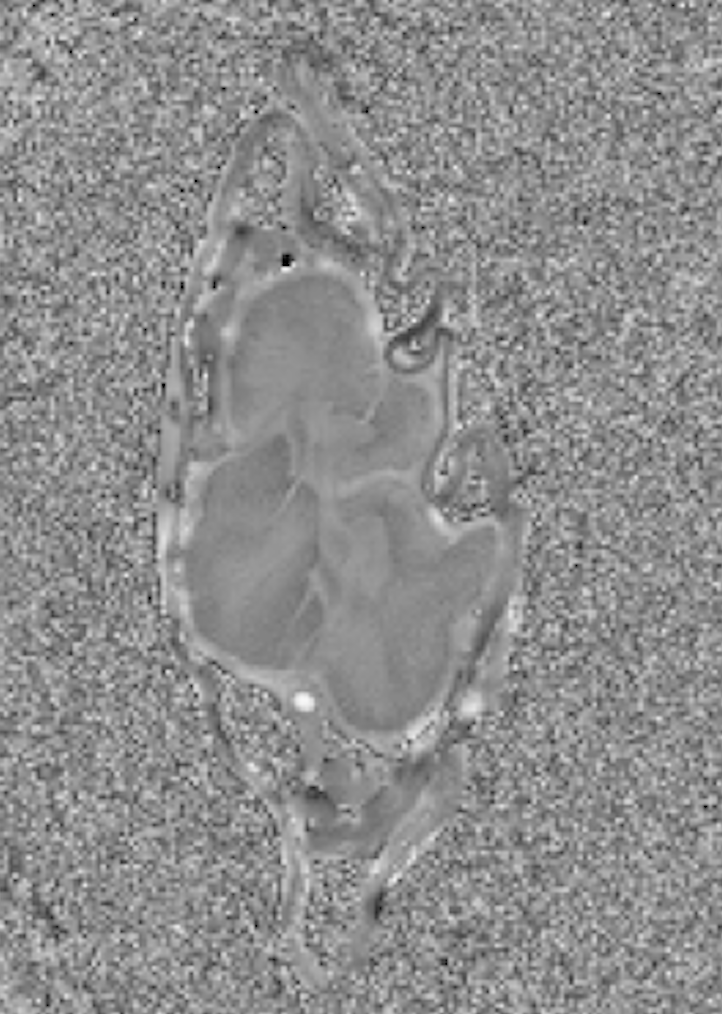}
        \caption{Original Phase}
        \label{fig:otsu_phase_orig}
    \end{subfigure}%
    \hspace{0.5cm}
    \begin{subfigure}[t]{0.4\columnwidth} % <--- Changed [b] to [t]
        \centering
        \includegraphics[width=\textwidth]{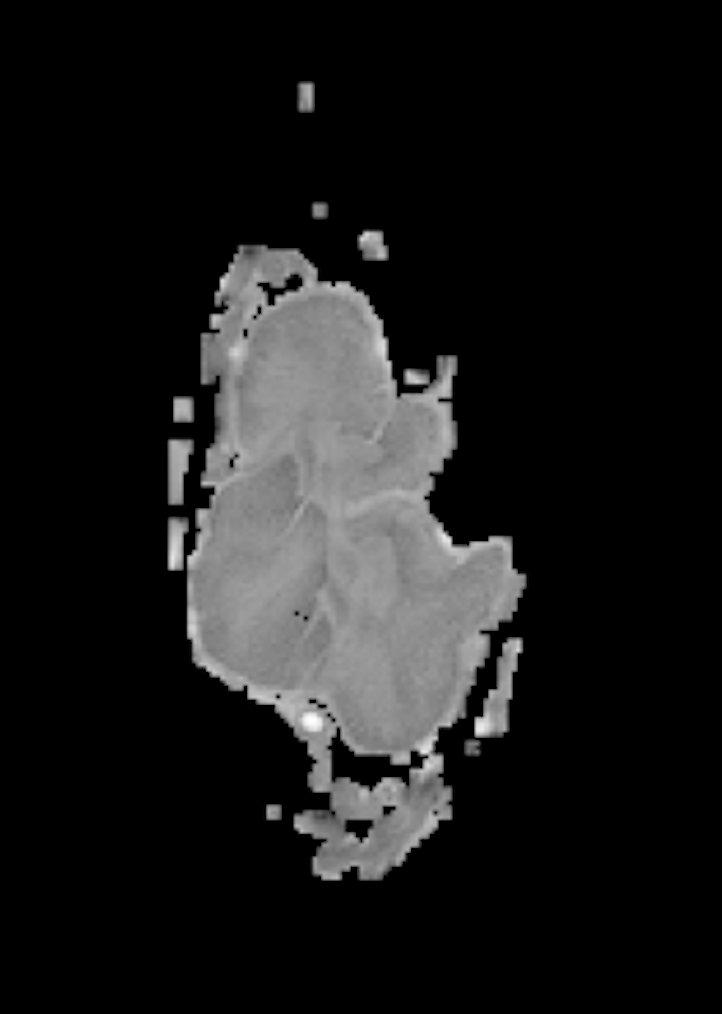}
        \caption{Otsu Masked Phase}
        \label{fig:otsu_phase}
    \end{subfigure}
    
    % --- Main Caption ---
    \caption{Illustration of the Otsu masking. The original magnitude (a), the masked magnitude (b), the original phase (c), and the masked phase (d).}
    \label{fig:otsu}
\end{figure}

\section{Background Masking: Hard Attention via Otsu Thresholding}
\label{app:otsu}

\subsubsection*{Motivation} 
High-resolution ex vivo MRI acquisitions are inherently noisy, often containing ghosting artifacts or flow artifacts in the non-tissue background air. Furthermore, the spinal cord occupies only a fraction of the total field of view, resulting in a massive class imbalance between ``background'' and ``tissue.'' We observed that in early experiments, the model would occasionally hallucinate lesions in the noisy background or struggle to converge because the vast majority of the loss signal came from empty space.

We believe that explicitly removing the background using a robust, unsupervised thresholding method could act as a ``hard attention'' mechanism as demonstrated by \cite{Tajbakhsh2020-kt}. By uniforming out non-tissue areas, we force the network to focus its entire capacity on the spinal cord ROI, removing false positives arising from background artifacts.

\subsubsection*{Methodology} 
We applied Otsu’s thresholding method to the T2* magnitude channel to generate a coarse binary mask. This mask was applied to both the magnitude and phase channels, ensuring that all background voxels were set to a uniform value (see Figure~\ref{fig:otsu}).

\subsubsection*{Results}
Table~\ref{tab:otsu_2D} shows the 2D results. Models trained without Otsu masking showed a degradation in performance (Dice Score ~1 percentage point decrease in the global score). Specifically, the total average Dice dropped from 0.766 (Base) to 0.758 (No Otsu). However, the Total Average HD95 distance slightly increased from 0.72 mm to 0.76 mm. To make a decision, we focused on the most challenging class: the Lesion GM. Otsu Masking improved the Dice Score by 2 percentage points and decreased the HD95 distance. Therefore we chose to keep the Otsu Masking variation.

\begin{table*}[t]
    \centering
    \caption{Otsu Masking 2D Experiment -- Dice and HD95 2D Results. Best results are highlighted in bold.}
    \label{tab:otsu_2D}
    \resizebox{\textwidth}{!}{%
        \begin{tabular}{lcccccccccc}
            \toprule
            & \multicolumn{2}{c}{\textbf{Healthy WM}} & \multicolumn{2}{c}{\textbf{Healthy GM}} & \multicolumn{2}{c}{\textbf{Lesion WM}} & \multicolumn{2}{c}{\textbf{Lesion GM}} & \multicolumn{2}{c}{\textbf{Total Average}} \\
            \cmidrule(lr){2-3} \cmidrule(lr){4-5} \cmidrule(lr){6-7} \cmidrule(lr){8-9} \cmidrule(lr){10-11}
            \textbf{Config} & \textbf{Dice} & \textbf{HD95} & \textbf{Dice} & \textbf{HD95} & \textbf{Dice} & \textbf{HD95} & \textbf{Dice} & \textbf{HD95} & \textbf{Dice} & \textbf{HD95} \\
            \midrule
            2D No Otsu & $0.847 \pm 0.087$ & $\mathbf{0.69 \pm 0.32}$ & $\mathbf{0.822 \pm 0.064}$ & $0.77 \pm 0.41$ & $0.720 \pm 0.042$ & $\mathbf{0.85 \pm 0.28}$ & $0.645 \pm 0.021$ & $0.59 \pm 0.26$ & $0.758 \pm 0.041$ & $\mathbf{0.72 \pm 0.30}$ \\
            2D Otsu Masking & $\mathbf{0.848 \pm 0.087}$ & $0.78 \pm 0.35$ & $0.819 \pm 0.067$ & $\mathbf{0.74 \pm 0.32}$ & $\mathbf{0.729 \pm 0.040}$ & $0.92 \pm 0.27$ & $\mathbf{0.665 \pm 0.021}$ & $\mathbf{0.58 \pm 0.28}$ & $\mathbf{0.766 \pm 0.041}$ & $0.76 \pm 0.28$ \\
            \bottomrule
        \end{tabular}%
    }
\end{table*}

Table~\ref{tab:otsu_3D} shows the 3D results. Unlike the 2D model, applying Otsu-based background masking was slightly detrimental to the 3D model's overall performance for both the total average and the challenging Lesion GM class. The total average Dice Score dropped marginally from 0.772 (No Otsu) to 0.764 (With Otsu Masking).

\begin{table*}[t]
    \centering
    \caption{Otsu Masking 3D Experiment -- Dice and HD95 3D Results. Best results are highlighted in bold.}
    \label{tab:otsu_3D}
    \resizebox{\textwidth}{!}{%
        \begin{tabular}{lcccccccccc}
            \toprule
            & \multicolumn{2}{c}{\textbf{Healthy WM}} & \multicolumn{2}{c}{\textbf{Healthy GM}} & \multicolumn{2}{c}{\textbf{Lesion WM}} & \multicolumn{2}{c}{\textbf{Lesion GM}} & \multicolumn{2}{c}{\textbf{Total Average}} \\
            \cmidrule(lr){2-3} \cmidrule(lr){4-5} \cmidrule(lr){6-7} \cmidrule(lr){8-9} \cmidrule(lr){10-11}
            \textbf{Config} & \textbf{Dice} & \textbf{HD95} & \textbf{Dice} & \textbf{HD95} & \textbf{Dice} & \textbf{HD95} & \textbf{Dice} & \textbf{HD95} & \textbf{Dice} & \textbf{HD95} \\
            \midrule
            3D No Otsu & $\mathbf{0.838 \pm 0.092}$ & $\mathbf{0.72 \pm 0.13}$ & $\mathbf{0.827 \pm 0.078}$ & $0.93 \pm 0.18$ & $\mathbf{0.750 \pm 0.083}$ & $1.38 \pm 0.92$ & $\mathbf{0.673 \pm 0.098}$ & $0.96 \pm 0.52$ & $\mathbf{0.772 \pm 0.079}$ & $1.00 \pm 0.35$ \\
            3D Otsu & $0.828 \pm 0.101$ & $0.72 \pm 0.14$ & $0.816 \pm 0.087$ & $\mathbf{0.75 \pm 0.24}$ & $0.744 \pm 0.079$ & $\mathbf{1.11 \pm 0.66}$ & $0.667 \pm 0.096$ & $\mathbf{0.75 \pm 0.42}$ & $0.764 \pm 0.087$ & $\mathbf{0.83 \pm 0.34}$ \\
            \bottomrule
        \end{tabular}%
    }
\end{table*}

\section{Optimization Strategies: AdamW vs. SGD}
\label{app:optimizer}

\subsubsection*{Motivation} 
The default configuration of nnU-Net relies on Stochastic Gradient Descent (SGD) with Nesterov momentum. This choice is grounded in extensive empirical benchmarking by \cite{Isensee2021-bc}, which demonstrated that SGD frequently yields superior generalization compared to adaptive solvers in standard biomedical segmentation tasks. However, the segmentation of multiple sclerosis lesions in the spinal cord presents a distinct set of challenges characterized by extreme class imbalance and sparse gradients from small pathological structures.

We hypothesized that the adaptive learning rates of AdamW (Adam with decoupled weight decay) might offer better convergence stability for these fine-grained targets than the fixed update schedules of SGD. While SGD is robust, AdamW’s ability to adjust the learning rate for each parameter individually can be advantageous in navigating the complex loss landscapes of small, difficult-to-detect lesions. To validate our deviation from the standard framework, we benchmarked the default SGD optimizer against the AdamW optimizer utilized in our 2D and 3D Base configurations.

\subsubsection*{Methodology} 
We trained a variation of the model using the standard SGD Optimizer while keeping all other parameters constant, allowing for a direct comparison with the AdamW-based Base models.

\begin{table*}[hb!]
    \centering
    \caption{Optimizer 2D Experiment -- Dice and HD95 2D Results. Best results are highlighted in bold.}
    \label{tab:opti_2D}
    \resizebox{\textwidth}{!}{%
        \begin{tabular}{lcccccccccc}
            \toprule
            & \multicolumn{2}{c}{\textbf{Healthy WM}} & \multicolumn{2}{c}{\textbf{Healthy GM}} & \multicolumn{2}{c}{\textbf{Lesion WM}} & \multicolumn{2}{c}{\textbf{Lesion GM}} & \multicolumn{2}{c}{\textbf{Total Average}} \\
            \cmidrule(lr){2-3} \cmidrule(lr){4-5} \cmidrule(lr){6-7} \cmidrule(lr){8-9} \cmidrule(lr){10-11}
            \textbf{Config} & \textbf{Dice} & \textbf{HD95} & \textbf{Dice} & \textbf{HD95} & \textbf{Dice} & \textbf{HD95} & \textbf{Dice} & \textbf{HD95} & \textbf{Dice} & \textbf{HD95} \\
            \midrule
            2D SGD & $0.838 \pm 0.088$ & $0.87 \pm 0.34$ & $0.804 \pm 0.058$ & $0.87 \pm 0.26$ & $0.687 \pm 0.072$ & $\mathbf{0.88 \pm 0.29}$ & $0.608 \pm 0.054$ & $0.62 \pm 0.27$ & $0.734 \pm 0.034$ & $0.81 \pm 0.25$ \\
            2D AdamW & $\mathbf{0.848 \pm 0.087}$ & $\mathbf{0.78 \pm 0.35}$ & $\mathbf{0.819 \pm 0.067}$ & $\mathbf{0.74 \pm 0.32}$ & $\mathbf{0.729 \pm 0.040}$ & $0.92 \pm 0.27$ & $\mathbf{0.665 \pm 0.021}$ & $\mathbf{0.58 \pm 0.28}$ & $\mathbf{0.766 \pm 0.041}$ & $\mathbf{0.76 \pm 0.28}$ \\
            \bottomrule
        \end{tabular}%
    }
\end{table*}

\begin{table*}[hb!]
    \centering
    \caption{Optimizer 3D Experiment -- Dice and HD95 3D Results. Best results are highlighted in bold.}
    \label{tab:opti_3D}
    \resizebox{\textwidth}{!}{%
        \begin{tabular}{lcccccccccc}
            \toprule
            & \multicolumn{2}{c}{\textbf{Healthy WM}} & \multicolumn{2}{c}{\textbf{Healthy GM}} & \multicolumn{2}{c}{\textbf{Lesion WM}} & \multicolumn{2}{c}{\textbf{Lesion GM}} & \multicolumn{2}{c}{\textbf{Total Average}} \\
            \cmidrule(lr){2-3} \cmidrule(lr){4-5} \cmidrule(lr){6-7} \cmidrule(lr){8-9} \cmidrule(lr){10-11}
            \textbf{Config} & \textbf{Dice} & \textbf{HD95} & \textbf{Dice} & \textbf{HD95} & \textbf{Dice} & \textbf{HD95} & \textbf{Dice} & \textbf{HD95} & \textbf{Dice} & \textbf{HD95} \\
            \midrule
            3D SGD & $0.834 \pm 0.091$ & $0.78 \pm 0.10$ & $0.826 \pm 0.078$ & $0.96 \pm 0.12$ & $0.733 \pm 0.076$ & $1.64 \pm 0.97$ & $0.667 \pm 0.087$ & $\mathbf{0.92 \pm 0.47}$ & $0.765 \pm 0.076$ & $1.07 \pm 0.34$ \\
            3D AdamW & $\mathbf{0.838 \pm 0.092}$ & $\mathbf{0.72 \pm 0.13}$ & $\mathbf{0.827 \pm 0.078}$ & $\mathbf{0.93 \pm 0.18}$ & $\mathbf{0.750 \pm 0.083}$ & $\mathbf{1.38 \pm 0.92}$ & $\mathbf{0.673 \pm 0.098}$ & $0.96 \pm 0.52$ & $\mathbf{0.772 \pm 0.079}$ & $\mathbf{1.00 \pm 0.35}$ \\
            \bottomrule
        \end{tabular}%
    }
\end{table*}

\subsubsection*{Results}

Table~\ref{tab:opti_2D} shows the 2D results. The use of AdamW yielded superior results across all semantic classes. This performance gain was particularly pronounced for the pathological structures, which represent the most challenging segmentation targets. Specifically, the transition to AdamW resulted in a Dice Score increase of approximately 4 percentage points for Lesion WM (0.729 vs. 0.687) and nearly 6 percentage points for Lesion GM (0.665 vs. 0.608) compared to the SGD counterpart.

Table~\ref{tab:opti_3D} shows the 3D results. Although the difference was less significant than in the 2D analysis, the 3D data showed that the AdamW optimizer (Base) surpassed the standard SGD implementation in overall performance metrics. Specifically, AdamW achieved a total average Dice Score of 0.772, better than SGD's 0.765. This advantage was most evident in the segmentation of Lesion WM, where AdamW's Dice Score was 0.750 compared to 0.733 for SGD.

\section{Synergy Analysis: Limits of 2D Regularization}
\label{app:synergy}
Having identified two robust independent strategies (Strong Spatial Augmentation (Strategy 2) and Lesion-Focused Soft Labeling (Soft Loss 2)) for the 2D Teacher Model, the logical next step was to evaluate their synergistic potential. We hypothesized that combining shape-invariant learning with boundary uncertainty modeling would yield a superior model.

\begin{table*}
    \centering
    \caption{2D Synergy Analysis -- Aug2 and Soft2 incompatibility -- Dice and HD95 2D Results. Best results are highlighted in bold.}
    \label{tab:2D_synergy}
    \resizebox{\textwidth}{!}{%
        \begin{tabular}{lcccccccccc}
            \toprule
            & \multicolumn{2}{c}{\textbf{Healthy WM}} & \multicolumn{2}{c}{\textbf{Healthy GM}} & \multicolumn{2}{c}{\textbf{Lesion WM}} & \multicolumn{2}{c}{\textbf{Lesion GM}} & \multicolumn{2}{c}{\textbf{Total Average}} \\
            \cmidrule(lr){2-3} \cmidrule(lr){4-5} \cmidrule(lr){6-7} \cmidrule(lr){8-9} \cmidrule(lr){10-11}
            \textbf{Config} & \textbf{Dice} & \textbf{HD95} & \textbf{Dice} & \textbf{HD95} & \textbf{Dice} & \textbf{HD95} & \textbf{Dice} & \textbf{HD95} & \textbf{Dice} & \textbf{HD95} \\
            \midrule
            2D Base (Aug1 + Hard) & $0.848 \pm 0.087$ & $\mathbf{0.78 \pm 0.35}$ & $0.819 \pm 0.067$ & $0.74 \pm 0.32$ & $0.729 \pm 0.040$ & $0.92 \pm 0.27$ & $0.665 \pm 0.021$ & $0.58 \pm 0.28$ & $0.766 \pm 0.041$ & $0.76 \pm 0.28$ \\
            2D Aug2 + Soft2 & $0.843 \pm 0.095$ & $0.85 \pm 0.50$ & $0.820 \pm 0.069$ & $0.78 \pm 0.40$ & $0.723 \pm 0.071$ & $\mathbf{0.82 \pm 0.28}$ & $0.654 \pm 0.024$ & $0.61 \pm 0.30$ & $0.760 \pm 0.037$ & $0.77 \pm 0.34$ \\
            2D Aug2 & $\mathbf{0.851 \pm 0.090}$ & $0.85 \pm 0.48$ & $0.825 \pm 0.064$ & $\mathbf{0.73 \pm 0.32}$ & $0.743 \pm 0.033$ & $0.84 \pm 0.34$ & $0.667 \pm 0.021$ & $\mathbf{0.53 \pm 0.23}$ & $0.772 \pm 0.042$ & $\mathbf{0.74 \pm 0.33}$ \\
            2D Soft2 & $0.846 \pm 0.094$ & $0.80 \pm 0.42$ & $\mathbf{0.828 \pm 0.064}$ & $0.74 \pm 0.31$ & $\mathbf{0.744 \pm 0.019}$ & $0.91 \pm 0.40$ & $\mathbf{0.672 \pm 0.008}$ & $0.56 \pm 0.29$ & $\mathbf{0.772 \pm 0.041}$ & $0.75 \pm 0.33$ \\
            \bottomrule
        \end{tabular}%
    }
\end{table*}

Counter-intuitively, integrating both techniques simultaneously proved detrimental as shown in Table~\ref{tab:2D_synergy}. Rather than compounding the benefits, the combination degraded performance compared to utilizing either strategy in isolation. As detailed in the table below, the Aug2 + Soft2 approach achieved a total average Dice of 0.760, a notable regression from the 0.772 achieved by the Soft Labeling or Spatial Augmentation models alone.

We attribute this phenomenon to destructive over-regularization. Spatial augmentation introduces significant variance to the input data (input noise), while soft labeling introduces uncertainty to the ground truth (target noise). When combined, these mechanisms likely reduced the training signal-to-noise ratio below a critical threshold, preventing the model from converging on fine-grained anatomical details.
Consequently, we selected Soft Segmentation 2 as our definitive Winning 2D Configuration to advance to the pseudo-labeling stage, citing its superior balance of average Dice Scores and stability on the challenging Lesion GM class. 

\section{Biophysical Basis of Phase Contrast in Spinal Cord MRI }
\label{app:biophysics}

While T2*-magnitude imaging relies on relaxation times, phase imaging derives contrast from the magnetic susceptibility of the underlying tissues. This provides a distinct advantage in the spinal cord, where the chemical composition of gray and white matter creates opposing frequency shifts \citep{Kearney2015-th}:
\begin{itemize}
    \item Gray matter (Paramagnetic): The phase contrast in gray matter is dominantly driven by the accumulation of non-heme iron, primarily stored in the protein ferritin \citep{Hulst2011-lo}. Although spinal cord gray matter contains lower iron concentrations than deep brain nuclei, the paramagnetic properties of iron attract the magnetic field, causing a positive frequency shift that renders the tissue distinct from the surrounding white matter.
    \item White matter (Diamagnetic): In contrast, the magnetic susceptibility of white matter is predominantly determined by myelin \citep{Geurts2008-ec}. Myelin is rich in lipids and proteins which are diamagnetic; they create a magnetic field opposite to the applied field, resulting in a negative frequency shift \citep{De_Leener2017-um, Hesamian2019-gn}.
\end{itemize}

This effect is particularly strong in the spinal cord due to the dense, coherent axonal tracts of the white matter columns. Consequently, even when a lesion alters the T2*w signal (magnitude), the underlying susceptibility difference often remains detectable in the Phase data, allowing for precise anatomical localization.

\end{document}